# Transverse transport of microswimmers in oscillatory channel flows

**Raghav Ram**[1] and **Ashwin Ramachandran**[1]†

[1] School of Mechanical Engineering, Purdue University, West Lafayette, Indiana 47907, USA

† Corresponding author: **ashwinrc@purdue.edu**

Motile microorganisms inhabit oscillatory flows encountered in physiological and engineering systems, yet the influence of flow unsteadiness on their shear-induced preferential concentration remains less understood. We investigate the transverse transport of elongated microswimmers in oscillatory pressure-driven channel flow using complementary Langevin simulations and two- and one-dimensional Fokker-Planck models. Excellent agreement between the particle-based and continuum descriptions across governing parameters establishes the fidelity of the Fokker-Planck formulation for microswimmer transport in oscillatory shear flows. Our analysis demonstrates that oscillatory forcing fundamentally modifies classical steady-flow shear-trapping. We find that increasing the Womersley number $Wo$ reduces centerline depletion by confining oscillatory shear to thinner near-wall regions. Contrastingly, we observe that increasing the frequency ratio $\beta$, of microswimmer rotational diffusion rate and flow oscillation frequency, promotes the development of sustained orientational anisotropy and leads to a saturating increase in centerline depletion. In the weak-swimming limit, we derive a hierarchy of coupled orientational moments, enabling analytical solutions at arbitrary temporal harmonic order. The asymptotic solutions establish a universal transfer law showing that, for a given microswimmer shape, the normalized leading-order orientational response depends solely on the frequency ratio $\beta$. Reconstruction of the orientational distribution yields closed-form expressions for the microswimmer concentration profile and depletion index, $I_D$. Our asymptotic solution recovers the steady-flow weak-shear scaling of $I_D$ with a flow Peclet number, $Pe_f$, of $I_D \propto Pe_f^2$, and reveals that oscillatory forcing attenuates this response by a factor $16\beta^2/(1 + 16\beta^2)$. Together, these results provide a theoretical framework for predicting transverse transport of swimming microorganisms in oscillatory channel flows.

## 1. Introduction

The transport of motile microorganisms and synthetic microswimmers in confined flows has attracted considerable attention because of its importance in biological, environmental and engineering systems, including microbial ecology, biomedical microfluidics, filtration, wastewater treatment and bioreactor design. Unlike passive tracers, self-propelled particles continuously swim while simultaneously undergoing hydrodynamic reorientation and rotational diffusion, giving rise to transport phenomena unique to active matter (Schnitzer, 1993; Bearon, 2003; Saintillan & Shelley, 2013; Elgeti et al., 2015; Bechinger et al., 2016). The coupling between orientational dynamics and cross-stream swimming gives rise to preferential concentration, including wall accumulation and centerline depletion, making the prediction of microorganism transport an important problem in fluid mechanics (Rusconi et al., 2014; Rusconi & Stocker, 2015).

A particularly important mechanism governing the redistribution of elongated swimming microorganisms is shear trapping. In pressure-driven channel flows, fluid shear preferentially aligns microswimmers with the streamwise direction, thereby reducing their wall-normal swimming velocity and increasing their residence time in high-shear regions near the channel walls. This mechanism gives rise to depletion of microswimmers near the channel centerline and accumulation near the walls. Rusconi et al., 2014 demonstrated this phenomenon experimentally and theoretically, showing that preferential concentration results from the competition between shear-induced alignment, active swimming, and rotational diffusion. Related experiments have demonstrated shear-induced orientational dynamics and spatial heterogeneity in motile phytoplankton (Barry et al., 2015), while kinetic descriptions have reproduced the trapping of slender bacteria in high-shear regions of channel flows (Bearon & Hazel, 2015; Ezhilan & Saintillan, 2015). Complementary studies based on deterministic and stochastic descriptions have further characterized microswimmer trajectories and transport in pressure-driven flows (Ganesh et al., 2023; Junot et al., 2019; Zöttl & Stark, 2012, 2013). Recent theoretical work has also developed continuum descriptions of shear-induced migration in pressure-driven channel flow, providing asymptotic transport equations for dilute suspensions of active Brownian particles (Vennamneni et al., 2020; Fung et al., 2022). Shear trapping can also couple microswimmer spatial distributions with the transport of dissolved chemical species to generate chemical concentration gradients that can influence microswimmer physiology and gene expression (Ramachandran et al., 2024). Together, these works have established shear trapping as one of the fundamental transport mechanisms governing elongated swimming microorganisms in steady channel flows.

More recently, attention has turned to the transport of active particles in oscillatory channel flows. Temporal forcing has been shown to substantially modify the long-time dispersion of active particles, providing new opportunities for controlling transport in microfluidic systems (Caldag & Bees, 2025; Chakraborty et al., 2026; Wang et al., 2025). These studies primarily examine Taylor dispersion and streamwise transport, demonstrating that oscillatory flows can either enhance or suppress effective dispersion depending on the swimmer dynamics and flow parameters. However, the influence of oscillatory forcing on shear trapping and cross-stream redistribution has received comparatively little attention. In particular, continuum descriptions capable of predicting concentration profiles and analytical scaling laws for oscillatory shear remain largely unavailable.

Oscillatory pressure-driven flows, in particular, are encountered in numerous physiological and engineering applications, including pulsatile blood flow, respiratory transport, oscillatory microfluidic devices and periodically driven bioreactors. Unlike steady flows, oscillatory shear introduces an additional timescale associated with the forcing frequency. Whether microorganisms can develop the orientational anisotropy responsible for shear trapping therefore depends not only on the magnitude of the imposed shear but also on its competition with the oscillation frequency, rotational diffusion, and swimming. As we formally establish in **Section 3.2**, these competing processes in oscillatory flow are naturally characterized by four dimensionless groups, namely, the flow Peclet number $Pe_f$, measuring the strength of shear rate relative to rotational diffusion; the swim Peclet number $Pe_c$, comparing swimming speed with rotational diffusion; the Womersley number $Wo$, characterizing the penetration of oscillatory forcing across the channel; and the frequency ratio $\beta$, comparing the timescales of rotational diffusion and flow oscillation. The resulting transport behavior is therefore expected to depend on the coupled influence of these dimensionless groups rather than on any single physical parameter. Despite the practical importance of oscillatory flows, the mechanisms governing preferential concentration under oscillatory forcing and the associated scaling laws have not yet been systematically established.

The motion of swimming microorganisms is commonly described using either stochastic particle models or continuum kinetic theories (Fung et al., 2025; Saintillan & Shelley, 2013). Langevin formulations resolve the trajectories of individual microswimmers by combining deterministic swimming, hydrodynamic reorientation and stochastic rotational diffusion, while the corresponding Fokker-Planck equation describes the evolution of the joint probability density of microswimmer position and orientation. Although the continuum formulation provides an efficient framework for predicting population-level transport, analytical solutions remain challenging because of the strong coupling between orientational dynamics and wall-normal migration. Reduced kinetic descriptions therefore provide an attractive means of obtaining analytical insight into the physical mechanisms governing microorganism transport.

The present work combines particle-based simulations, continuum modelling, and Fourier harmonic analysis to investigate the transport of elongated swimming microorganisms in oscillatory pressure-driven channel flow. We first formulate a Langevin model describing the stochastic motion of individual microswimmers and use it to identify the physical mechanisms governing preferential concentration in oscillatory flows. We then derive the corresponding two-dimensional Fokker–Planck equation governing the joint wall-normal position–orientation probability density and verify its predictions against the Langevin simulations over a broad range of flow conditions. The verified continuum formulation is subsequently used to systematically characterize the competing influences of shear-induced alignment, rotational diffusion, wall-normal swimming, and oscillatory forcing through four governing dimensionless groups. Finally, under the assumption of weak swimming, we derive a reduced one-dimensional Fokker-Planck equation governing the local orientational distribution. The orientational dynamics are analyzed using a Fourier harmonic representation, which yields a hierarchy of coupled evolution equations for the Fourier moments of the orientational distribution. Exploiting this hierarchy, we obtain asymptotic solutions for the leading harmonics, derive a universal transfer law governing the orientational response to oscillatory shear, and obtain analytical expressions for the period-averaged microswimmer concentration profile and depletion index.

The remainder of the paper is organized as follows. Section 2 presents the oscillatory channel flow together with the Langevin model, the two-dimensional Fokker-Planck formulation and the reduced one-dimensional Fokker-Planck equation. Section 3 first investigates the mechanisms governing preferential concentration using Langevin simulations, then validates the continuum formulation against the particle-based model, systematically examines the influence of the governing dimensionless parameters using the two-dimensional Fokker-Planck equation, and finally develops a Fourier harmonic analysis of the reduced one-dimensional model that yields a hierarchy of orientational moments and analytical expressions for the orientational response, concentration profile, and depletion index, which are validated against numerical results.

## 2. Theory

We consider the transport of motile microswimmers confined between two parallel plates (or walls) and subjected to an oscillatory pressure-driven flow (**Figure 1**). This section consists of four parts. First, the imposed oscillatory flow is outlined. Second, the motion of individual microswimmers is described using Langevin equations for position and orientation. Third, the corresponding two-dimensional Fokker-Planck equation for the joint position-orientation probability density, is outlined. Thereafter, a reduced one-dimensional Fokker-Planck equation for the local orientational distribution at a fixed wall-normal position is derived. Finally, we derive closed-form solutions for the orientational harmonics and scaling laws for integral measures of the period-averaged concentration distribution using asymptotic analysis of the reduced Fokker–Planck equation.

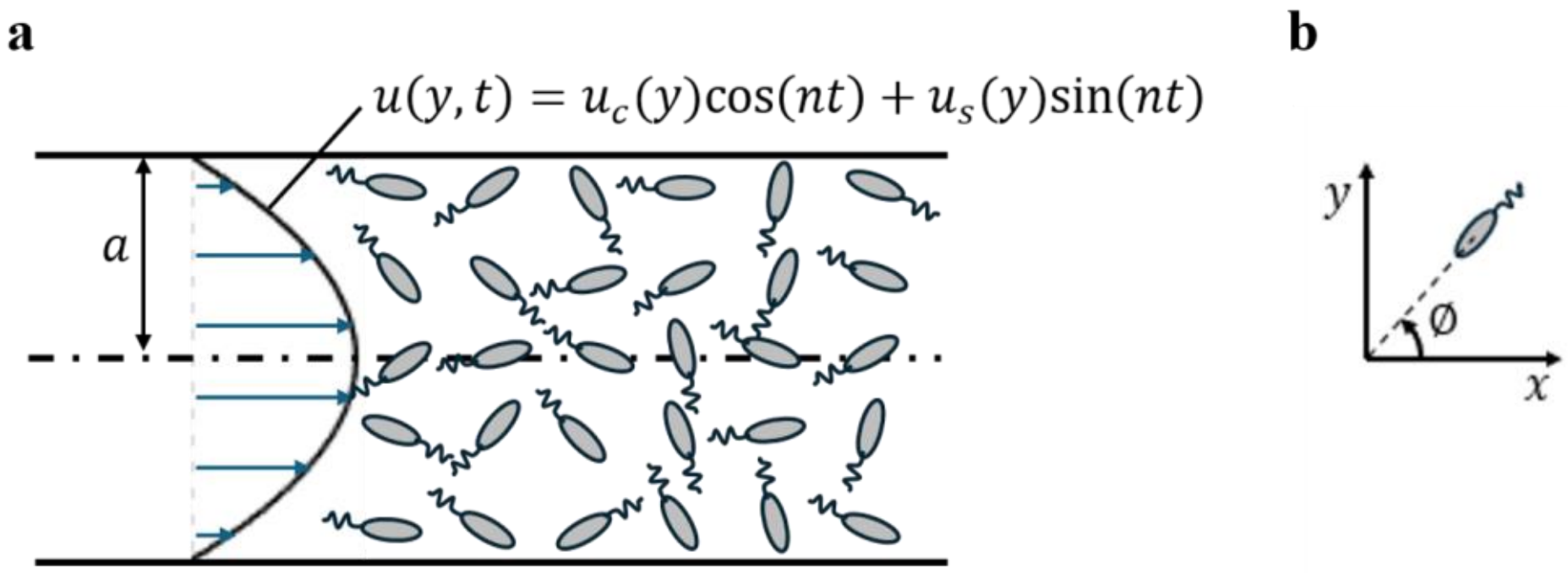

**Figure 1**. Schematic of microswimmer transport in oscillatory pressure-driven channel flow. (a) A dilute suspension of microswimmers confined between two parallel plates with channel half-width $a$, subjected to the imposed flow $\boldsymbol{u} = u(y,t)\ \boldsymbol{e_x}$. (b) The $x - y$ coordinate system and definition of the microswimmer orientation angle $\emptyset$, measured counterclockwise from the positive $x$-axis to the swimmer body axis.

### 2.1. Oscillatory channel flow

We consider a dilute suspension of microswimmers confined between two flat parallel plates located at $y = -a$ and $y = a$ (**Figure 1**). In the dilute limit, the influence of the microswimmers on the fluid flow is assumed to be negligible, allowing the flow field to be determined independently of the microswimmer dynamics. We consider a streamwise-homogeneous microswimmer distribution, with no dependence on $x$**.** The fluid velocity is likewise independent of $x$, and is described by a one-dimensional flow, $\boldsymbol{u} = u(y,t)\ \boldsymbol{e_x}$, of a fluid of density $\rho$ and kinematic viscosity $\nu$, driven by an oscillatory pressure gradient (Womersley, 1955), given by,

$$\frac{\partial p}{\partial x} = -A\cos nt, \tag{2.1}$$

where $\boldsymbol{e}_x$ is the unit vector in the streamwise direction, $a$ is the channel half-width, $A$ is the amplitude of the pressure gradient, and $n$ is the angular frequency of oscillation. The solution for velocity is given by (Kurzweg, 1985; Landau & Lifshitz, 1959; Loudon & Tordesillas, 1998),

$$\begin{aligned} u(y,t) = \frac{A}{n\rho\gamma}\{ &[\sinh f_1(y)\sin f_2(y) + \sinh f_2(y)\sin f_1(y)]\cos nt \\ &+ [\gamma - \cosh f_1(y)\cos f_2(y) - \cosh f_2(y)\cos f_1(y)]\sin nt \}, \end{aligned} \tag{2.2}$$

where,

$$f_1(y) = \left(1+\frac{y}{a}\right)\frac{Wo}{\sqrt{2}}, \qquad f_2(y) = \left(1-\frac{y}{a}\right)\frac{Wo}{\sqrt{2}}, \qquad \gamma = \cosh\left(\sqrt{2}Wo\right) + \cos\left(\sqrt{2}Wo\right).$$

Here, $Wo = a\sqrt{n/\nu}$ is the Womersley number that quantifies the relative importance of unsteady inertial effects and viscous diffusion in oscillatory internal flows. Equivalently, $Wo^2$ represents the ratio of the viscous diffusion timescale across the channel to the oscillation timescale.

We non-dimensionalize the wall-normal coordinate, time, and velocity according to,

$$\hat{y} = \frac{y}{a}, \qquad \tau = nt, \qquad \hat{u} = \frac{u}{aS}, \tag{2.3}$$

where $S = Aa/(\nu\rho)$ is the maximum wall shear rate in the quasi-steady ($Wo \to 0$) limit, which we use as the characteristic scale for shear rate in this study.

The dimensionless velocity can then be written as follows:

$$\begin{aligned} \hat{u}(\hat{y},\tau;Wo) = \frac{1}{Wo^2\gamma}\{ &\left[\sinh \hat{f}_1(\hat{y})\sin \hat{f}_2(\hat{y}) + \sinh \hat{f}_2(\hat{y})\sin \hat{f}_1(\hat{y})\right]\cos\tau \\ &+ \left[\gamma - \cosh \hat{f}_1(\hat{y})\cos \hat{f}_2(\hat{y}) - \cosh \hat{f}_2(\hat{y})\cos \hat{f}_1(\hat{y})\right]\sin\tau \}, \end{aligned} \tag{2.4}$$

where $\hat{f}_1(\hat{y}) = (1+\hat{y})\,Wo/\sqrt{2}$, and $\hat{f}_2(\hat{y}) = (1-\hat{y})\,Wo/\sqrt{2}$.

The flow profiles corresponding to $Wo = 0.1$ , 1, and 10 at eight different timepoints within an oscillation cycle are shown in **Figure 2**. At $Wo = 0.1$, the instantaneous velocity profiles remain nearly parabolic throughout the oscillation period, consistent with the quasi-steady limit. As $Wo$ increases, the profiles progressively depart from the parabolic form and become increasingly plug-like in the channel interior, with the strongest velocity gradients localized near the walls at $Wo =$ 10. These changes in the spatial distribution of the shear rate help to interpret the microswimmer concentration distributions in **Section 3.2.1**.

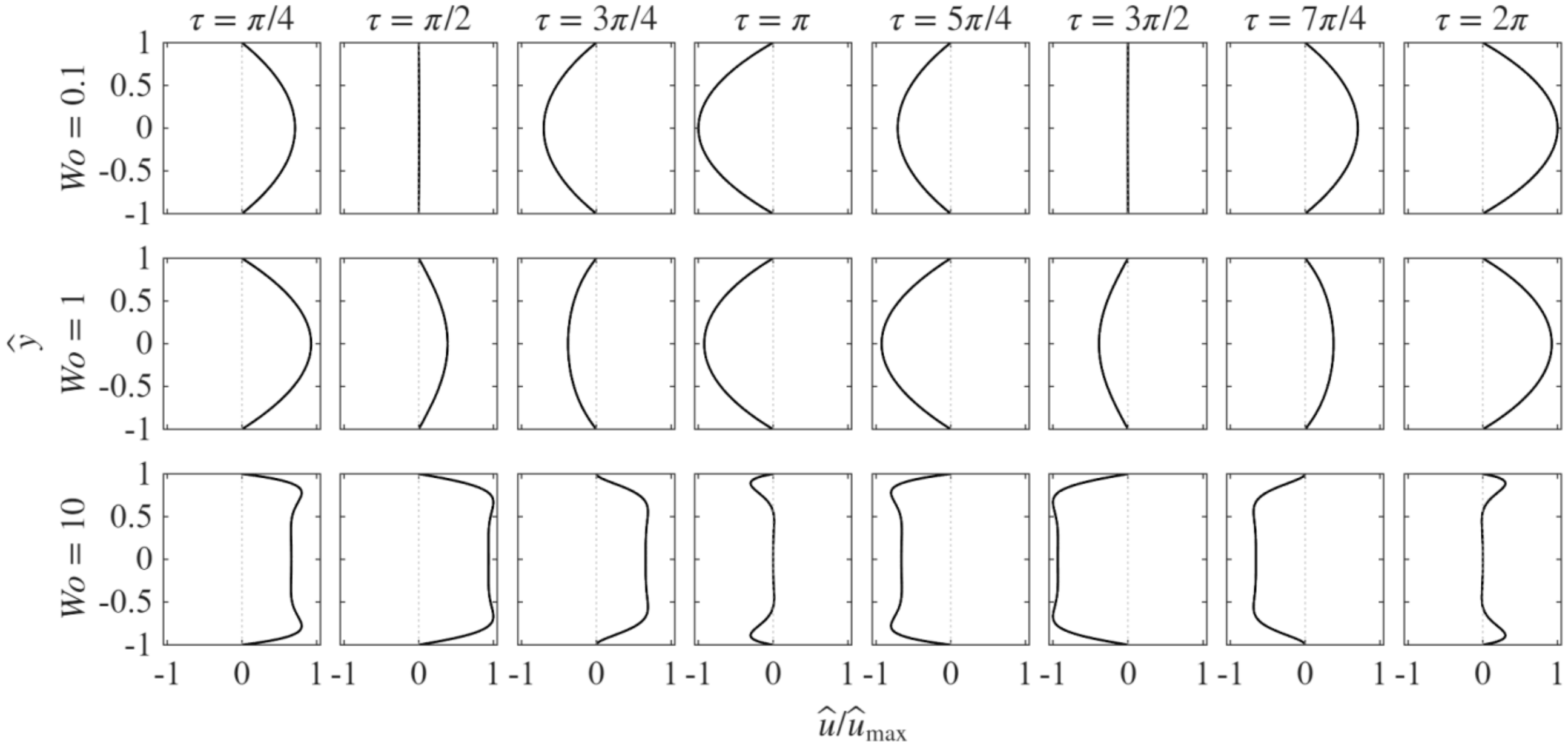

**Figure 2**. Effect of Womersley number $Wo$ on the unsteady flow profile normalized by the maximum velocity $\hat{u}_{\max}$ within an oscillation period at timepoints $\tau = \pi/4\,,\pi/2\,,3\pi/4\,,\pi,5\pi/4\,,3\pi/2\,,7\pi/4\,,2\pi$, where $2\pi$ is the non-dimensional oscillation time period.

### 2.2. Langevin Model

The position and orientation of an individual microswimmer are described by its streamwise coordinate $x$, wall-normal coordinate $y$, and its orientation angle $\emptyset$ measured relative to the streamwise direction. The Langevin equations governing the motion of a microswimmer having an aspect ratio $q$, swimming speed $V$ and rotational diffusivity $D_R$ are given by,

$$\frac{dx}{dt} = -V\cos\emptyset + u(y,t),$$

$$\frac{dy}{dt} = -V\sin\emptyset, \tag{2.5}$$

$$\frac{d\emptyset}{dt} = \frac{1}{2}\frac{\partial u}{\partial y}[\alpha\cos(2\emptyset) - 1] + \sqrt{2D_R}\,\eta(t),$$

where the deterministic shear-induced rotation is described by Jeffery's equation (Jeffery, 1922) with the particle aspect ratio entering through the shape factor, $\alpha = (q^2 - 1)/(q^2 + 1)$, $\eta(t)$ is a Gaussian white noise process describing rotational Brownian motion, and $u(y,t)$ is the flow field outlined in **Section 2.1**. We model the microswimmers as prolate ellipsoids with aspect ratio $q = 10$ throughout this study.

For numerical integration, the equations may be discretized as follows.

$$\Delta x = -\Delta t\, V \cos \emptyset + \Delta t\, u(y,t),$$

$$\Delta y = -\Delta t\, V \sin \emptyset, \tag{2.6}$$

$$\Delta \emptyset = \frac{1}{2} \Delta t \frac{\partial u}{\partial y} [\alpha \cos(2\emptyset) - 1] + \sqrt{2 D_R \Delta t}\, \eta_n,$$

where $\eta_n \sim \mathcal{N}(0,1)$ is an independent standard normal random variable sampled at each time step.

The dynamics of this problem are governed by five timescales, namely, $2\pi/n$, $1/S$, $a/V$, $1/D_R$, and $a^2/\nu$, and a Buckingham Π analysis yields the following four dimensionless groups,

$$\frac{S}{D_R} = Pe_f, \qquad \frac{V}{a D_R} = Pe_c, \qquad a\sqrt{\frac{n}{\nu}} = Wo, \qquad \frac{D_R}{n} = \beta, \tag{2.7}$$

where, $Pe_f$ is the flow Peclet number, $Pe_c$ is the microswimmer Peclet number, $Wo$ is the Womersley number, and $\beta$ is the ratio of the rotational diffusion rate to the oscillation frequency.

Using the characteristic length, time, and velocity scales defined in **Section 2.1**, Eqs. 2.6 may be expressed in dimensionless form as follows:

$$\Delta \hat{x} = -Pe_c \beta\, \Delta\tau \cos \emptyset + Pe_f \beta\, \Delta\tau\, \hat{u}(\hat{y}, \tau; Wo),$$

$$\Delta \hat{y} = -Pe_c \beta\, \Delta\tau \sin \emptyset, \tag{2.8}$$

$$\Delta \emptyset = \frac{1}{2} Pe_f \beta\, \Delta\tau \frac{\partial \hat{u}}{\partial \hat{y}} [\alpha \cos(2\emptyset) - 1] + \sqrt{2\beta \Delta\tau}\, \eta_n.$$

It is worth noting that the dynamics in the streamwise coordinate $\hat{x}$ is decoupled from the $(\hat{y}, \emptyset)$ dynamics and does not affect the wall-normal concentration profiles of interest in this study. Consequently, the $\hat{x}$-coordinate is omitted from the remainder of the analysis. The Langevin model results we present in **Section 3** are obtained by integrating the equations of motion (Eqs. 2.8) using a fourth-order Runge–Kutta method for the deterministic terms, and rotational Brownian motion is included by adding the stochastic angular increment $\sqrt{2\beta\Delta\tau}\, \eta_n$ at each time step. The time step for numerical integration is chosen based on the condition, $\Delta\tau = 0.01 \times \min\{2\pi, 1/(Pe_f \beta), 1/(Pe_c \beta), 1/\beta\}$, and the trajectories of $5 \times 10^5$ microswimmers, initialized at random positions $\hat{y} \in [-1,1]$ and orientations $\emptyset \in [-\pi, \pi]$, are integrated from $\tau = 0$ to $\tau = 10 \times \max\{2\pi, 1/(Pe_f \beta), 1/(Pe_c \beta), 1/\beta\}$. Specularly reflective boundary conditions are imposed at the channel walls, $\hat{y} = \pm 1$, i.e., when a microswimmer crosses a wall, its position is reflected back into the domain and its orientation is reversed according to $\phi \to -\phi$, reversing the wall-normal component of the swimming velocity while preserving the streamwise component.

### 2.3. Fokker-Planck formulation

Denoting $p(y, \emptyset, t)$ as the joint probability density of finding a microswimmer at position $y$ with orientation $\emptyset$ at time $t$, the evolution of $p(y, \emptyset, t)$ is given by the Fokker-Planck equation (FPE) corresponding to the Langevin equations of motion. The FPE corresponding to the $(\dot{y}, \dot{\emptyset})$ equations (Eq. 2.5) is given by,

$$\frac{\partial p}{\partial t} = V \sin \emptyset \frac{\partial p}{\partial y} - \frac{1}{2} \frac{\partial u}{\partial y} \frac{\partial}{\partial \emptyset}([\alpha \cos(2\emptyset) - 1]p) + D_R \frac{\partial^2 p}{\partial \emptyset^2}. \tag{2.9}$$

In terms of dimensionless variables, the FPE for the evolution of $\hat{p}(\hat{y}, \emptyset, \tau)$ can be written as,

$$\frac{\partial \hat{p}}{\partial \tau} = Pe_c \beta \sin \emptyset \frac{\partial \hat{p}}{\partial \hat{y}} - \frac{1}{2} Pe_f \beta \frac{\partial \hat{u}}{\partial \hat{y}} \frac{\partial}{\partial \emptyset}([\alpha \cos(2\emptyset) - 1]\hat{p}) + \beta \frac{\partial^2 \hat{p}}{\partial \emptyset^2}, \tag{2.10a}$$

where,

$$\frac{\partial \hat{u}}{\partial \hat{y}} = \frac{1}{Wo}\{F_1(\hat{y}; Wo) \cos \tau + F_2(\hat{y}; Wo) \sin \tau\},$$

$$\begin{aligned} F_1(\hat{y}; Wo) = \frac{1}{\sqrt{2}\,\gamma} \big[&- \sinh \hat{f}_1(\hat{y}) \cos \hat{f}_2(\hat{y}) + \cosh \hat{f}_1(\hat{y}) \sin \hat{f}_2(\hat{y}) \\ &+ \sinh \hat{f}_2(\hat{y}) \cos \hat{f}_1(\hat{y}) - \cosh \hat{f}_2(\hat{y}) \sin \hat{f}_1(\hat{y})\big], \\ F_2(\hat{y}; Wo) = \frac{1}{\sqrt{2}\,\gamma} \big[&- \cosh \hat{f}_1(\hat{y}) \sin \hat{f}_2(\hat{y}) - \sinh \hat{f}_1(\hat{y}) \cos \hat{f}_2(\hat{y}) \\ &+ \cosh \hat{f}_2(\hat{y}) \sin \hat{f}_1(\hat{y}) + \sinh \hat{f}_2(\hat{y}) \cos \hat{f}_1(\hat{y})\big]. \end{aligned} \tag{2.10b}$$

The definitions of $\hat{f}_1$, $\hat{f}_2$ and $\gamma$ can be found in **Section 2.1**. Equivalently, Eq. 2.10a can be rewritten as follows.

$$\begin{aligned} \frac{\partial^2 \hat{p}}{\partial \emptyset^2} - \frac{Pe_f}{2Wo}\{F_1(\hat{y}; Wo) \cos \tau + F_2(\hat{y}; Wo) \sin \tau\} \frac{\partial}{\partial \emptyset}([\alpha \cos(2\emptyset) - 1]\hat{p}) \\ + Pe_c \sin \emptyset \frac{\partial \hat{p}}{\partial \hat{y}} = \frac{1}{\beta} \frac{\partial \hat{p}}{\partial \tau}. \end{aligned} \tag{2.10c}$$

In the FPE, the wall-normal flux term represents cross-stream swimming transport, the orientation-space advection term describes shear-induced reorientation, the diffusion term accounts for rotational randomization, and the unsteady term captures the temporal response of the microswimmer distribution to the oscillatory flow. The interplay of these mechanisms governs the evolution of the joint position–orientation distribution and the resulting preferential concentration patterns.

We here solve the FPE numerically for $\hat{p}(\hat{y}, \emptyset, \tau)$, subject to periodic boundary conditions in orientation, $\hat{p}(\hat{y}, -\pi, \tau) = \hat{p}(\hat{y}, \pi, \tau)$ and $\partial_\emptyset \hat{p}(\hat{y}, -\pi, \tau) = \partial_\emptyset \hat{p}(\hat{y}, \pi, \tau)$, and specularly reflective boundary conditions at the channel walls, $\hat{p}(\pm 1, \emptyset, \tau) = \hat{p}(\pm 1, -\emptyset, \tau)$. We use an initial condition corresponding to a uniform distribution in position and orientation, $\hat{p}(\hat{y}, \emptyset, 0) = 1/(4\pi)$.

Integrating the joint probability density over orientation space yields the probability density of finding a microswimmer at position $\hat{y}$. This quantity is equivalent to the normalized microswimmer concentration and is given by,

$$C(\hat{y}, \tau) = \int_{-\pi}^{\pi} \hat{p}(\hat{y}, \emptyset, \tau)\ d\emptyset. \tag{2.11}$$

Integrating the FPE equation (2.10) over $-\pi \le \emptyset \le \pi$ eliminates the orientation-space advection and diffusion terms owing to the periodic boundary conditions in orientation, resulting in the following evolution equation for normalized microswimmer concentration.

$$\frac{\partial C}{\partial \tau} - Pe_c \beta \frac{\partial}{\partial \hat{y}} \left[ \int_{-\pi}^{\pi} \sin \emptyset\ \hat{p}(\hat{y}, \emptyset, \tau)\ d\emptyset \right] = 0 \tag{2.12}$$

Equation (2.12) may be interpreted as a continuity equation for microswimmer concentration, with the wall-normal swimming flux, $J = -Pe_c \beta \int_{-\pi}^{\pi} \sin \emptyset\ \hat{p}(\hat{y}, \emptyset, \tau)\ d\emptyset$, determined by the first orientational moment, $\int_{-\pi}^{\pi} \sin \emptyset\ \hat{p}\ d\emptyset$. Accordingly, concentration changes arise only through spatial gradients in the wall-normal swimming flux, while a spatially uniform flux leaves the concentration unchanged.

**2.4. One-dimensional Fokker-Planck equation**

While the two-dimensional Fokker–Planck equation discussed in **Section 2.3** provides a complete description of the joint evolution of microswimmer position and orientation, additional insight can be obtained by examining the orientational dynamics at a fixed wall-normal location. To this end, we consider the orientational probability density $p(\emptyset, t; y)$, describing the distribution of the local microswimmer orientation $\emptyset$ at time $t$, at a fixed position $y$. With $y$ being treated as a parameter, the cross-stream swimming transport term in the 2D-FPE (Eq. 2.9) is omitted, and the resulting one-dimensional Fokker-Planck equation (1D-FPE) is given by,

$$\frac{\partial p}{\partial t} = -\frac{1}{2} \frac{\partial u}{\partial y} \frac{\partial}{\partial \emptyset} ([\alpha \cos(2\emptyset) - 1] p) + D_R \frac{\partial^2 p}{\partial \emptyset^2}. \tag{2.13}$$

In terms of dimensionless variables, the 1D-FPE for the evolution of $\hat{p}(\emptyset, \tau; \hat{y})$ can be written as,

$$\frac{\partial \hat{p}}{\partial \tau} = -\frac{1}{2} Pe_f \beta \frac{\partial \hat{u}}{\partial \hat{y}} \frac{\partial}{\partial \emptyset} ([\alpha \cos(2\emptyset) - 1] \hat{p}) + \beta \frac{\partial^2 \hat{p}}{\partial \emptyset^2}. \tag{2.14a}$$

The definition of $\partial \hat{u} / \partial \hat{y}$ can be found in **Section 2.3**. Equivalently, Eq. 2.14a can be rewritten as follows.

$$\frac{\partial^2 \hat{p}}{\partial \emptyset^2} - \frac{Pe_f}{2Wo} \{F_1(\hat{y}; Wo) \cos \tau + F_2(\hat{y}; Wo) \sin \tau\} \frac{\partial}{\partial \emptyset} ([\alpha \cos(2\emptyset) - 1] \hat{p}) = \frac{1}{\beta} \frac{\partial \hat{p}}{\partial \tau}. \tag{2.14b}$$

The 1D-FPE isolates the effects of local shear-induced reorientation and rotational diffusion on the orientational distribution. For a given $\hat{y}$, we solve the 1D-FPE numerically for $\hat{p}(\emptyset,\tau;\hat{y})$, subject to periodic boundary conditions in orientation, $\hat{p}(-\pi,\tau;\hat{y}) = \hat{p}(\pi,\tau;\hat{y})$ and $\partial_{\emptyset}\hat{p}(-\pi,\tau;\hat{y}) = \partial_{\emptyset}\hat{p}(\pi,\tau;\hat{y})$. Similar to 2D-FPE, we use an initial condition corresponding to a uniform distribution in orientation, $\hat{p}(\emptyset,0;\hat{y}) = 1/(2\pi)$. We note that the 1D-FPE is not directly comparable to the Langevin model because it neglects cross-stream swimming transport, and therefore, its applicability is restricted to the limit $Pe_c \ll 1$. However, as we show in **Section 3**, the reduced dimensionality of the 1D-FPE makes it amenable to perturbation and harmonic analyses, enabling the derivation of asymptotic expressions for normalized microswimmer concentration and depletion index.

## 3. Results and Discussion

This section investigates the mechanisms governing microswimmer transport and redistribution in oscillatory flow through a sequence of complementary modeling approaches. We begin with Langevin simulations, which provide a particle-level description of microswimmer trajectories and serve as the reference solution throughout this study. The influence of the four governing dimensional parameters, flow frequency ($n$), shear rate ($S$), swimming speed ($V$) and rotational diffusivity ($D_R$), is first analyzed through period-averaged concentration profiles and joint position–orientation probability density functions. The solutions of the Fokker–Planck formulation (2D-FPE) are then verified against Langevin simulations to demonstrate the ability of the continuum framework to reproduce the dynamics predicted by the agent-based model. Following this verification, the 2D-FPE is employed to systematically investigate the effects of the governing dimensionless groups, $Wo$, $Pe_f$, $Pe_c$, $\beta$, on preferential concentration. Next, we consider the reduced one-dimensional Fokker–Planck equation (1D-FPE), derived under the assumption of weak swimming. Its predictions are compared with those of the Langevin model to establish the parameter regime over which the reduced model remains accurate. Finally, we obtain asymptotic solutions from the 1D-FPE through perturbation and harmonic analyses. We derive analytical expressions for the normalized period-averaged concentration profile and depletion index and validate them against numerical results.

### 3.1. Langevin model

Langevin simulations are employed to investigate the influence of the governing dimensional parameters on microswimmer transport in oscillatory flow. The results are presented in terms of normalized period-averaged concentration profiles, $\langle C(\hat{y})\rangle$, and period-averaged joint position–orientation probability density functions, $\langle \hat{p}(\hat{y},\emptyset)\rangle$, which respectively characterize the spatial distribution of microswimmers and the underlying orientational dynamics responsible for transport. The effects of flow frequency ($n$), shear rate ($S$), swimming speed ($V$) and rotational diffusivity ($D_R$) are examined to identify the mechanisms governing preferential concentration in oscillatory flows.

### 3.1.1 Frequency-induced suppression of preferential concentration

Previous studies of microswimmers in steady shear flows have demonstrated that fluid shear can give rise to preferential concentration, characterized by depletion of microswimmers near the channel centerline and accumulation near the walls (Rusconi et al. 2014). We first examine whether this mechanism persists in oscillatory flows and how it is modified by the temporal variation of the shear field. **Figure 3** shows the effect of flow frequency on microswimmer transport, keeping the other parameters constant at $S = 100\ \mathrm{s}^{-1}$, $V = 50\ \mu\mathrm{m\ s}^{-1}$ and $D_R = 1\ \mathrm{rad}^2\ \mathrm{s}^{-1}$.

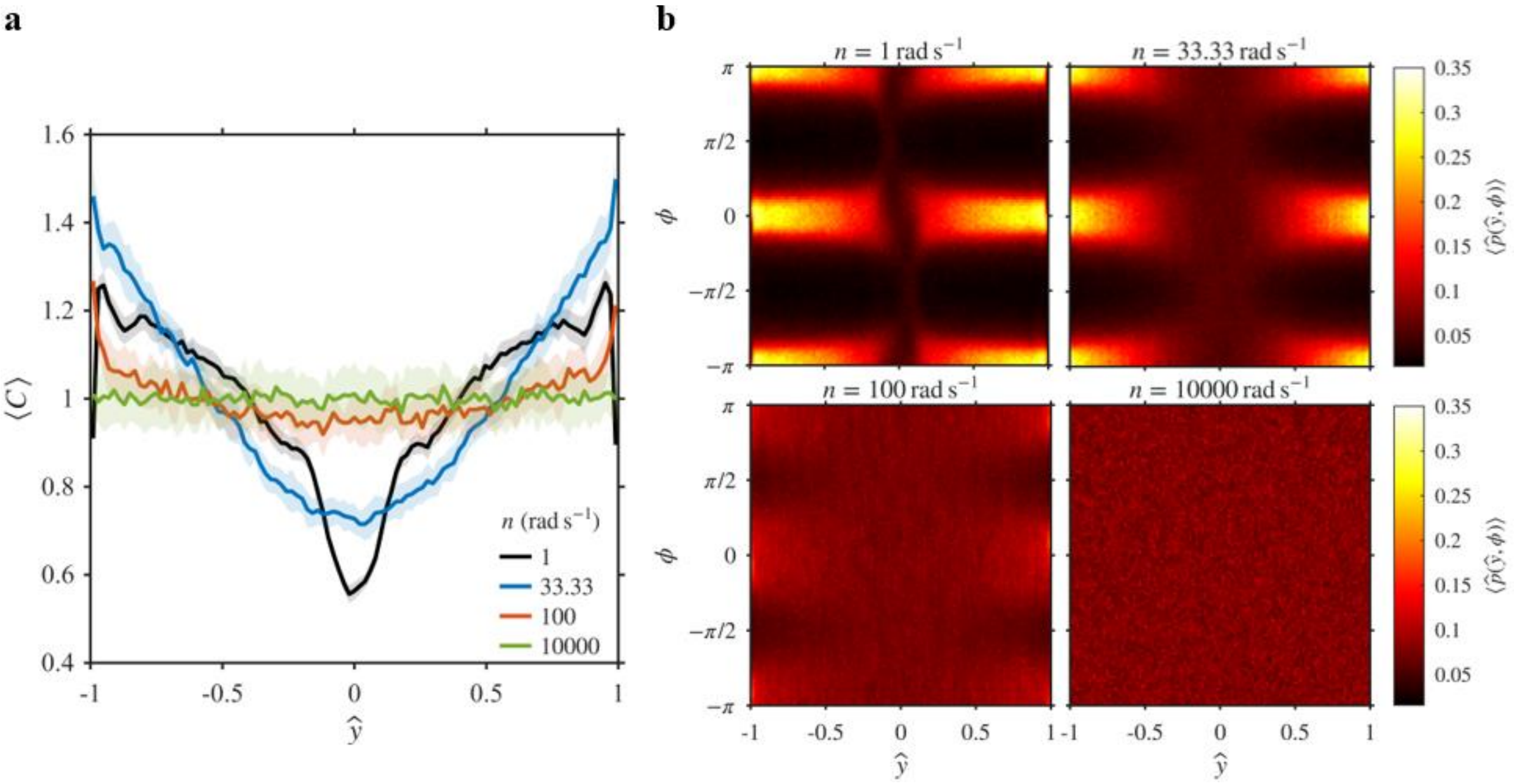

**Figure 3**. Effect of oscillation frequency on period-averaged (a) concentration $\langle C \rangle$ versus $\hat{y}$, and (b) joint-probability density function $\langle \hat{p} \rangle$ versus $\emptyset$ and $\hat{y}$, for $S = 100\ \mathrm{s}^{-1}$, $V = 50\ \mu\mathrm{m\ s}^{-1}$ and $D_R = 1\ \mathrm{rad}^2\ \mathrm{s}^{-1}$. Shaded regions in (a) denote one standard deviation about the mean period-averaged concentration over 20 realizations.

The period-averaged concentration profiles exhibit a strong dependence on oscillation frequency (**Figure 3a**). At low frequencies of $n = 1\ \mathrm{rad\ s}^{-1}$ and $n = 33.33\ \mathrm{rad\ s}^{-1}$, we observe substantial centerline depletion and wall accumulation, consistent with the shear-induced depletion reported in steady flows. The associated probability density functions (**Figure 3b**) exhibit strong bands near $\emptyset = 0$ and $\emptyset = \pm\pi$, indicating that microswimmers are preferentially aligned with the streamwise direction due to their elongated shape, consistent with the underlying mechanism for shear-trapping (Rusconi et al., 2014; Vennamneni et al., 2020). This orientation distribution varies across the channel width and generates a non-uniform wall-normal swimming flux (Eq. 2.12), resulting in microswimmer redistribution and preferential concentration. As the oscillation frequency increases to $n = 100\ \mathrm{rad\ s}^{-1}$ and $n = 10000\ \mathrm{rad\ s}^{-1}$, the concentration profile progressively approaches a uniform distribution and the corresponding orientational distribution becomes increasingly diffuse and homogeneous. These results suggest that rapid oscillations inhibit the development of the sustained orientational states required for shear-induced migration. Additionally, while low-frequency forcing allows microswimmers sufficient time to respond to the instantaneous shear field and undergo significant cross-stream redistribution, high-frequency forcing repeatedly reverses the direction of the shear before substantial transverse migration can occur. These effects together result in a suppression of preferential concentration distribution, i.e., shear-trapping, as flow oscillation frequency increases.

We next examine how the instantaneous concentration distribution varies within a flow oscillation period. **Figure 4** shows the variation of instantaneous concentration profiles at $\tau = \pi/4, \pi/2, 3\pi/4, \pi, 5\pi/4, 3\pi/2, 7\pi/4,$ and $2\pi$ within a flow oscillation period, at $n = 1$ rad s$^{-1}$, $S = 100$ s$^{-1}$, $V = 50$ $\mu$m s$^{-1}$ and $D_R = 1$ rad$^2$ s$^{-1}$.

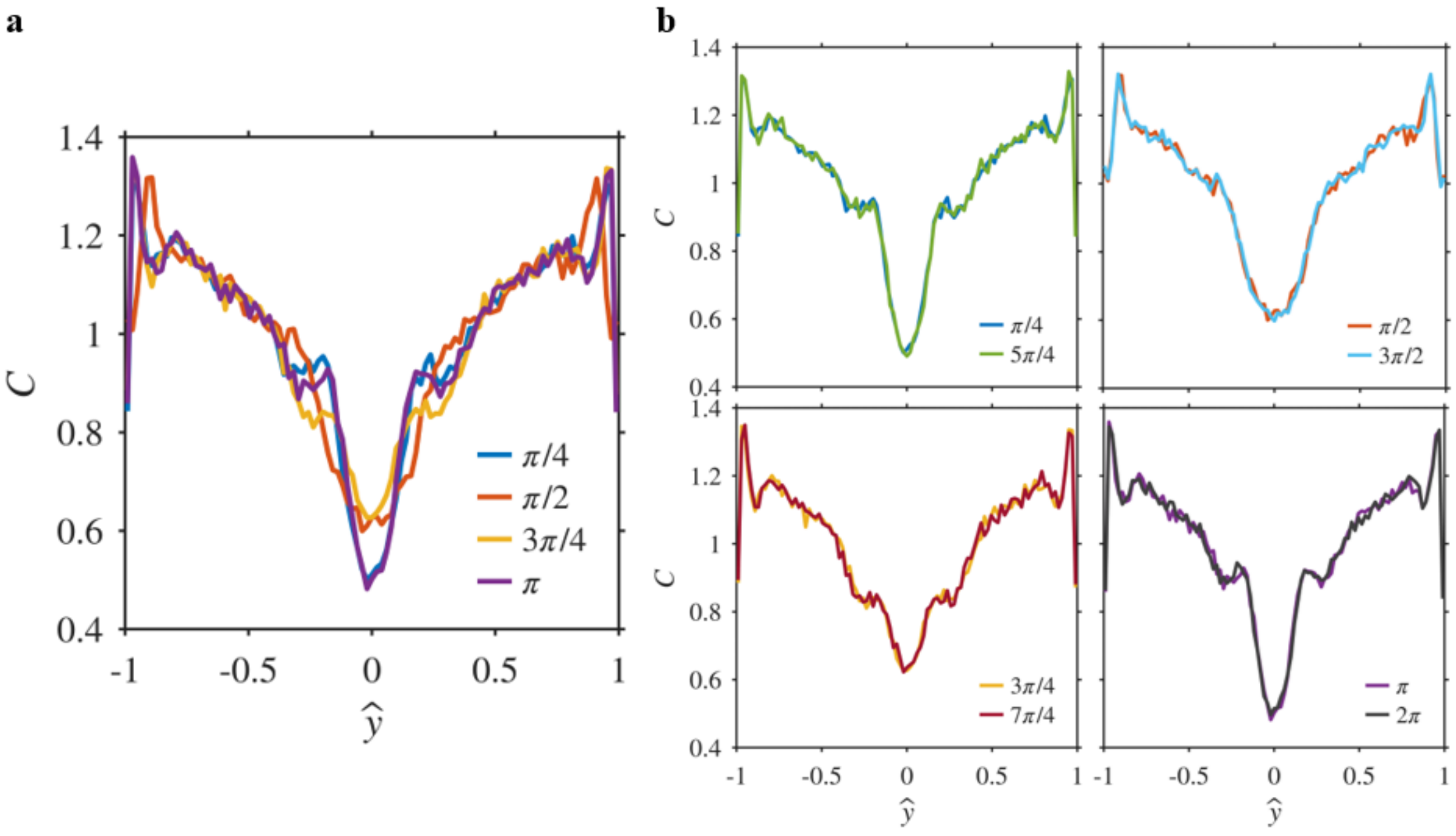

**Figure 4**. Concentration $C$ versus $\hat{y}$ (a) at $\tau = \pi/4, \pi/2, 3\pi/4,$ and $\pi$, within a flow oscillation period, and (b) at timepoints $\tau$ and $\tau + \pi$ within a flow oscillation period, for $n = 1$ rad s$^{-1}$, $S = 100$ s$^{-1}$, $V = 50$ $\mu$m s$^{-1}$ and $D_R = 1$ rad$^2$ s$^{-1}$.

From **Figure 4**, we observe that the concentration remains symmetric about the channel centerline $\hat{y} = 0$, within the statistical fluctuations inherent to Langevin simulations. The concentration also exhibits modest temporal variation over the course of an oscillation period. A clear half-period symmetry is evident, with concentration profiles separated by $\Delta\tau = \pi$ collapsing almost exactly onto one another (**Figure 4b**). We rationalize these observations using the equivalent 2D Fokker-Planck equation (Eq. 2.10c), which reveals the underlying symmetries of the problem. First, the equation is invariant under the reflection transformation $(\hat{y}, \emptyset, \tau) \rightarrow (-\hat{y}, -\emptyset, \tau)$, implying that the long-time periodic solution (i.e., after the transient state due to initial conditions) satisfies $\hat{p}(\hat{y}, \emptyset, \tau) = \hat{p}(-\hat{y}, -\emptyset, \tau)$. Further, integrating $\hat{p}$ over orientation space yields $C(\hat{y}, \tau) = C(-\hat{y}, \tau)$, so the concentration profile is symmetric about the channel centerline $\hat{y} = 0$. Second, the 2D-FPE is invariant under the transformation $(\hat{y}, \emptyset, \tau) \rightarrow (\hat{y}, \pi - \emptyset, \tau + \pi)$, from which the long-time periodic solution satisfies $\hat{p}(\hat{y}, \emptyset, \tau) = \hat{p}(\hat{y}, \pi - \emptyset, \tau + \pi)$. Therefore, integrating $\hat{p}$ over orientation space immediately yields $C(\hat{y},\tau) = C(\hat{y},\tau+\pi)$, which shows that the concentration is $\pi$-periodic as seen in **Figure 4b** despite the imposed oscillatory flow having a period of $2\pi$. Finally, integrating the concentration continuity equation (Eq. 2.12) over one half-period gives $\partial_{\hat{y}}\left[\int_{\tau}^{\tau+\pi} J\,(\hat{y},\tau')\,\mathrm{d}\tau'\right] = 0$, which shows that the half-period-averaged wall-normal swimming flux is independent of $\hat{y}$. Since the specular wall boundary conditions, $\hat{p}(\pm 1, \emptyset, \tau) = \hat{p}(\pm 1, -\emptyset, \tau)$, imply $J(\pm 1, \tau) = 0$, it follows that $\int_{\tau}^{\tau+\pi} J(\hat{y},\tau')\,\mathrm{d}\tau' = 0$ throughout the channel. Therefore, in the long-time periodic state, although microswimmers undergo instantaneous wall-normal migration during each half-cycle, the net wall-normal swimming flux over half a period is zero across any cross-section.

**3.1.2 Shear-induced alignment and wall accumulation**

We here study the effect of oscillatory flow maximum shear rate on the microswimmer concentration distribution. **Figure 5** shows the period-averaged microswimmer concentration profiles for varying maximum shear rate values of $S = 1\ \mathrm{s}^{-1}$, $S = 10\ \mathrm{s}^{-1}$ and $S = 100\ \mathrm{s}^{-1}$, at two representative flow frequencies, $n = 1\ \mathrm{rad\ s}^{-1}$ and $n = 100\ \mathrm{rad\ s}^{-1}$, and $V = 50\ \mu\mathrm{m\ s}^{-1}$ and $D_R = 1\ \mathrm{rad}^2\ \mathrm{s}^{-1}$.

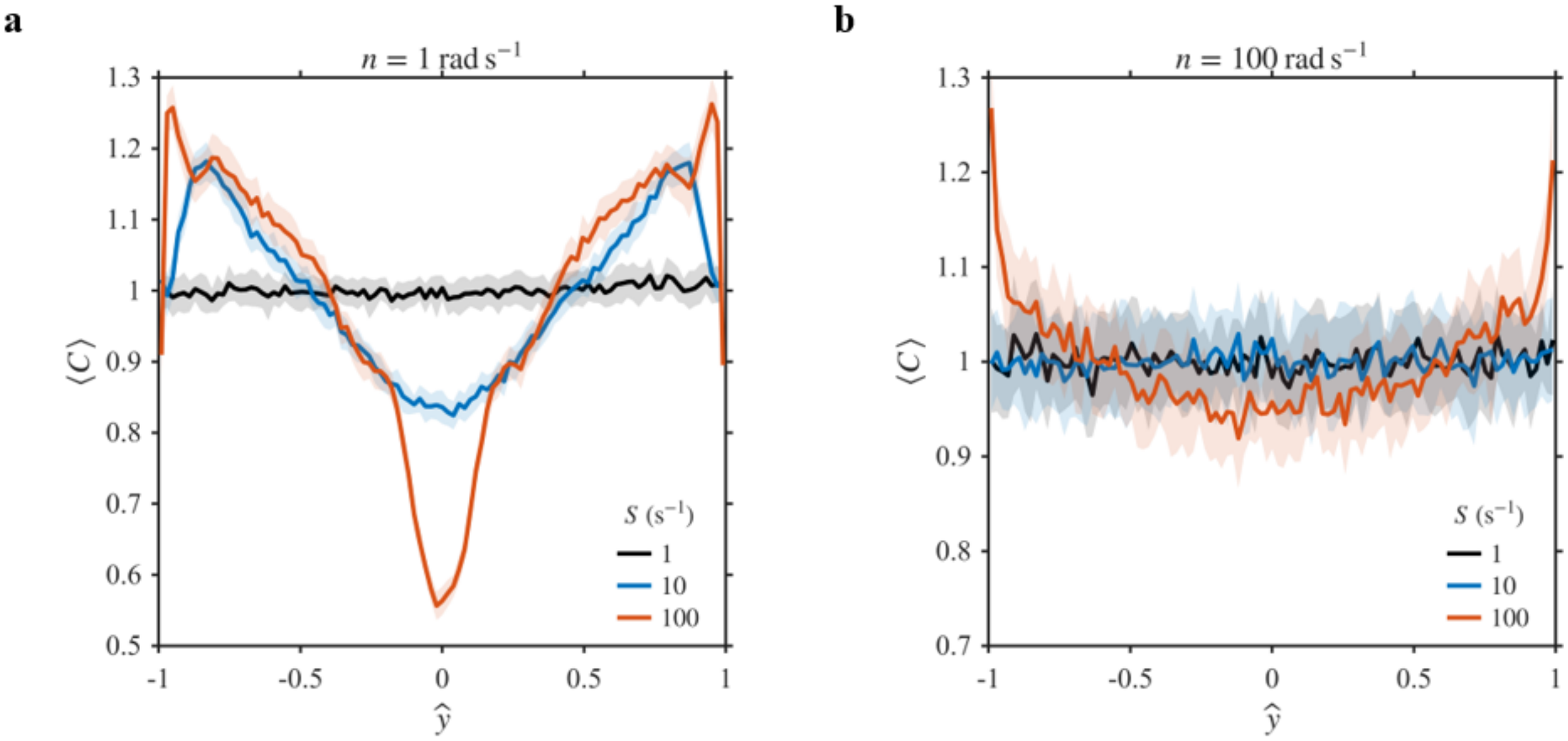

**Figure 5**. Effect of oscillatory flow maximum shear rate on period-averaged concentration $\langle C\rangle$ versus $\hat{y}$ for (a) $n = 1\ \mathrm{rad\ s}^{-1}$, and (b) $n = 100\ \mathrm{rad\ s}^{-1}$, for $V = 50\ \mu\mathrm{m\ s}^{-1}$ and $D_R = 1\ \mathrm{rad}^2\ \mathrm{s}^{-1}$. Shaded regions denote one standard deviation about the mean period-averaged concentration over 20 realizations.

At low oscillation frequency of $n = 1\ \mathrm{rad\ s}^{-1}$, increasing shear rate strongly enhances preferential concentration. For $S = 1\ \mathrm{s}^{-1}$, the concentration remains nearly uniform, whereas substantial centerline depletion develops for $S = 10\ \mathrm{s}^{-1}$ and this effect is pronounced at $S = 100\ \mathrm{s}^{-1}$. This behavior is consistent with the observations of Rusconi et al. (2014) and Vennamneni et al. (2020), who reported enhanced shear-induced depletion with increasing shear rate in steady flows, suggesting that the same shear trapping mechanism remains operative at low oscillation frequencies. At the high frequency limit, the influence of shear is considerably weaker. Concentration gradients appear only at $S = 100\ \mathrm{s}^{-1}$, while lower shear rates produce nearly uniform distributions. Therefore, we find that increasing the oscillatory flow shear rate increases centerline depletion and leads to preferential microswimmer concentration, although this effect is less pronounced at larger oscillation frequencies.

### 3.1.3 Competition between swimming and shear-induced trapping

We next analyze the effect of microswimmer swimming speed on the concentration distribution. **Figure 6** shows the period-averaged microswimmer concentration profiles for varying swimming speeds of $V = 0\ \mu\mathrm{m\ s}^{-1}$, $V = 5\ \mu\mathrm{m\ s}^{-1}$, $V = 50\ \mu\mathrm{m\ s}^{-1}$ and $V = 500\ \mu\mathrm{m\ s}^{-1}$, at two representative flow frequencies, $n = 1\ \mathrm{rad\ s}^{-1}$ and $n = 100\ \mathrm{rad\ s}^{-1}$, and $S = 100\ \mathrm{s}^{-1}$ and $D_R = 1\ \mathrm{rad}^2\ \mathrm{s}^{-1}$.

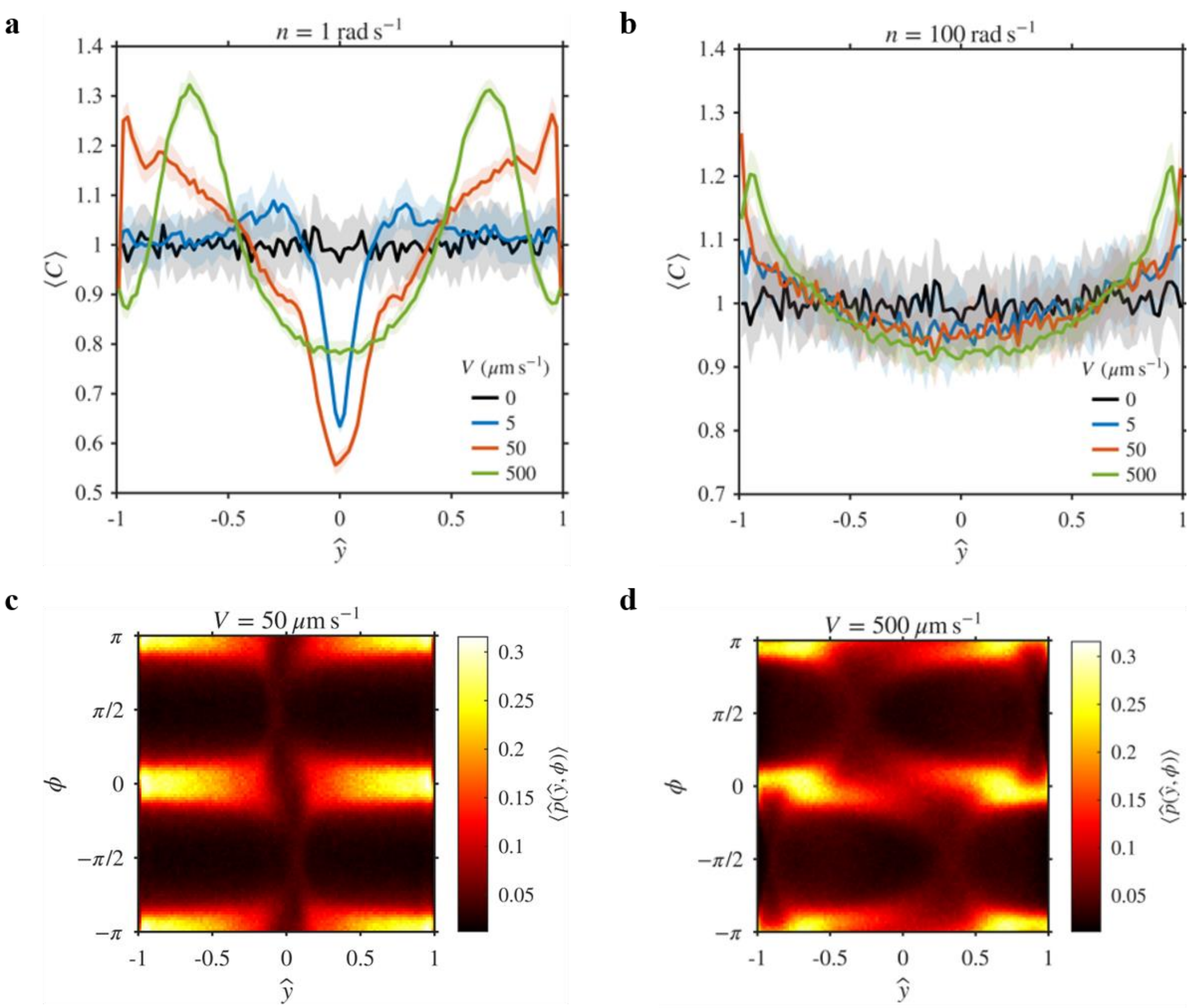

**Figure 6**. Effect of swimming speed on period-averaged concentration $\langle C \rangle$ versus $\hat{y}$ for (a) $n = 1\ \mathrm{rad\ s}^{-1}$, and (b) $n = 100\ \mathrm{rad\ s}^{-1}$, for $S = 100\ \mathrm{s}^{-1}$ and $D_R = 1\ \mathrm{rad}^2\ \mathrm{s}^{-1}$. Shaded regions denote one standard deviation about the mean period-averaged concentration over 20 realizations. Period-averaged joint-probability density function $\langle \hat{p} \rangle$ versus $\emptyset$ and $\hat{y}$, for (c) $V = 50\ \mu\mathrm{m\ s}^{-1}$, and (d) $V = 500\ \mu\mathrm{m\ s}^{-1}$, for $n = 1\ \mathrm{rad\ s}^{-1}$, $S = 100\ \mathrm{s}^{-1}$, and $D_R = 1\ \mathrm{rad}^2\ \mathrm{s}^{-1}$.

At $n = 1 \text{ rad s}^{-1}$, an increase in swimming speed initially strengthens centerline depletion, with the strongest depletion occurring at $V = 50\ \mu\text{m s}^{-1}$. A further increase in swimming speed to $V = 500\ \mu\text{m s}^{-1}$ weakens the centerline depletion and broadens the near-wall accumulation regions. We hypothesize this is because in this low oscillation frequency regime, increasing swimming speed initially enhances preferential concentration by strengthening the wall-normal transport associated with shear-induced orientational anisotropy. However, at very large swimming speeds ($V = 500\ \mu\text{m s}^{-1}$), microswimmers traverse the channel more rapidly than they can be reoriented by the local shear field, reducing the effectiveness of the shear-trapping mechanism. This is further reflected by considering the joint probability density function distribution, $\langle\hat{p}\rangle$. **Figure 6d** shows the period-averaged joint-probability density function for swimming speed $V = 500\ \mu\text{m s}^{-1}$, at $n = 1 \text{ rad s}^{-1}$, $S = 100 \text{ s}^{-1}$ and $D_R = 1 \text{ rad}^2 \text{ s}^{-1}$. From **Figure 6d**, we observe significant probabilities for intermediate orientations at $V = 500\ \mu\text{m s}^{-1}$, unlike the tight distributions seen in **Figure 6c** for $V = 50\ \mu\text{m s}^{-1}$, in addition to the pronounced bands near $\emptyset = 0$ and $\emptyset = \pm\pi$. We see that this reduction in streamwise orientational localization at $V = 500\ \mu\text{m s}^{-1}$ is accompanied by weaker centerline depletion (**Figure 6a**).

At a high oscillation frequency of $n = 100 \text{ rad s}^{-1}$, we observe a weaker dependence of period-averaged concentration profiles on swimming speed (**Figure 6b**). Although wall accumulation increases slightly with swimming speed, the overall concentration remains nearly uniform across the channel width. Similar to **Figure 5b**, we hypothesize that the observation of nearly uniform concentration profiles in **Figure 6b** is due to suppression of orientational ordering required to sustain strong migration in the presence of rapid flow oscillations.

**3.1.4 Rotational diffusion-mediated shear trapping**

We here study the effect of microswimmer rotational diffusivity on the concentration distribution. **Figure 7** shows the period-averaged microswimmer concentration profiles for varying rotational diffusivities of $D_R = 0 \text{ rad}^2 \text{ s}^{-1}$, $D_R = 1 \text{ rad}^2 \text{ s}^{-1}$, $D_R = 10 \text{ rad}^2 \text{ s}^{-1}$ and $D_R = 50 \text{ rad}^2 \text{ s}^{-1}$, at two representative flow frequencies, $n = 1 \text{ rad s}^{-1}$ and $n = 100 \text{ rad s}^{-1}$, and $S = 100 \text{ s}^{-1}$ and $V = 50\ \mu\text{m s}^{-1}$.

At a low flow oscillation frequency of $n = 1 \text{ rad s}^{-1}$, the most pronounced centerline depletion occurs for $D_R = 1 \text{ rad}^2 \text{ s}^{-1}$, while increasing rotational diffusivity from $1 \text{ rad}^2 \text{ s}^{-1}$ to $50 \text{ rad}^2 \text{ s}^{-1}$ progressively weakens the concentration gradients. For $D_R = 50 \text{ rad}^2 \text{ s}^{-1}$, the concentration becomes nearly uniform. When rotational diffusivity is small, microswimmers maintain their orientation for long periods and retain strong preferential orientations due to the flow shear rates (c.f. **Figure 3b**). This generates significant wall-normal swimming flux and therefore substantial concentration gradients. As $D_R$ increases, rotational diffusion randomizes microswimmer orientations and progressively disrupts the orientational structure. **Figure 7d** plots the period-averaged joint-probability density function for $D_R = 50 \text{ rad}^2 \text{ s}^{-1}$, for $n = 1 \text{ rad s}^{-1}$, $S = 100 \text{ s}^{-1}$, and $V = 50\ \mu\text{m s}^{-1}$. The reduction in orientational anisotropy as rotational diffusivity increases weakens the wall-normal swimming flux and suppresses preferential concentration.

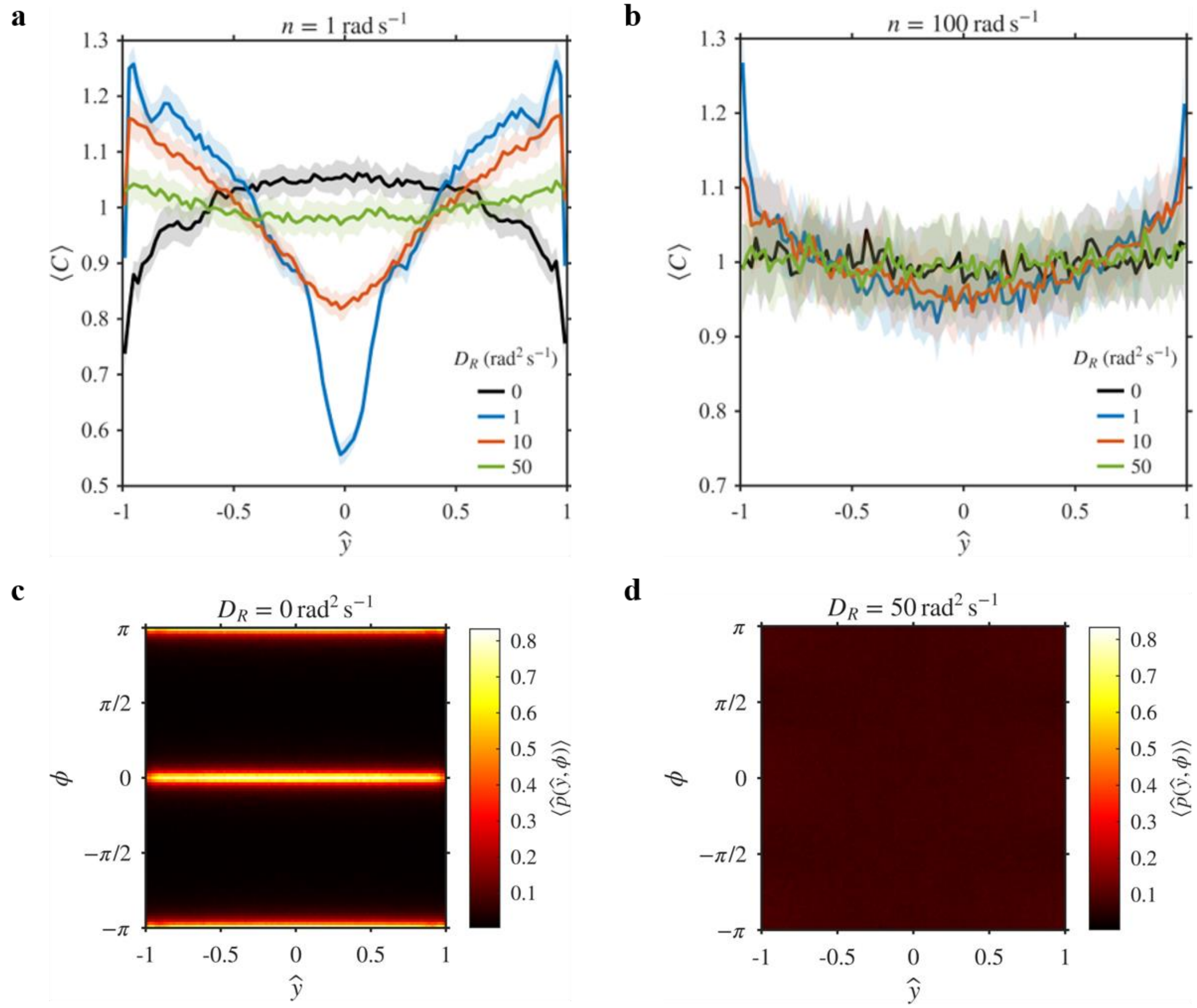

**Figure 7**. Effect of rotational diffusivity on period-averaged concentration $\langle C \rangle$ versus $\hat{y}$ for (a) $n = 1\ \text{rad s}^{-1}$, and (b) $n = 100\ \text{rad s}^{-1}$, for $S = 100\ \text{s}^{-1}$ and $V = 50\ \mu\text{m s}^{-1}$. Shaded regions denote one standard deviation about the mean period-averaged concentration over 20 realizations. Period-averaged joint-probability density function $\langle \hat{p} \rangle$ versus $\emptyset$ and $\hat{y}$, for (c) $D_R = 0\ \text{rad}^2\ \text{s}^{-1}$, and (d) $D_R = 50\ \text{rad}^2\ \text{s}^{-1}$, for $n = 1\ \text{rad s}^{-1}$, $S = 100\ \text{s}^{-1}$, and $V = 50\ \mu\text{m s}^{-1}$.

The behavior at $D_R = 0\ \text{rad}^2\ \text{s}^{-1}$ is particularly interesting. **Figure 7c** plots the period-averaged joint-probability density function for $D_R = 0\ \text{rad}^2\ \text{s}^{-1}$, for $n = 1\ \text{rad s}^{-1}$, $S = 100\ \text{s}^{-1}$, and $V = 50\ \mu\text{m s}^{-1}$. In this limit of zero rotational diffusion, we observe sharp peaks in the joint probability density function at $\emptyset = 0$ and $\emptyset = \pm\pi$. This is because in the absence of rotational diffusion, microswimmers follow deterministic trajectories in position–orientation phase space and become strongly localized in a limited set of orientations. As noted by Rusconi et al. (2014), rotational diffusion perturbs microswimmers away from their deterministic trajectories, which enables them to cross the separatrix in the $\hat{y} - \emptyset$ phase space (Zöttl & Stark, 2013) and access trajectories that sample the high-shear near-wall regions. Consequently, similar to steady flows, a small amount of rotational diffusion enhances wall accumulation and strengthens preferential concentration even for oscillatory flows. Lastly, **Figure 7b** shows that at a high oscillation frequency of $n = 100\ \text{rad s}^{-1}$, concentration gradients are substantially weaker for all values of rotational diffusivity, which reiterates the reduced ability of rapidly oscillating flows to generate the required persistent orientational anisotropy.

The results presented in this section were obtained by varying one (dimensional) parameter at a time while holding the remaining parameters fixed. This approach provides a basis for the range of microswimmer transport behaviors expected to be seen in oscillatory flows. While the qualitative mechanisms governing microswimmer transport and preferential concentration remain the same, the magnitude of the observed effects may depend on the particular combination of flow frequency, shear rate, swimming speed and rotational diffusivity. Therefore, next in **Section 3.2**, we expand on these observations and conduct a systematic exploration of the parameter space by considering the governing dimensionless parameters and the 2D-FPE framework.

### 3.2. Fokker-Planck formulation

An equivalent continuum Fokker-Planck formulation (2D-FPE) for the microswimmer-based Langevin description was derived in **Section 2.3**. In this section, we first assess the consistency of the 2D-FPE with the microswimmer-based description by comparing its predictions with the Langevin simulations described in **Section 3.1**. **Figure 8** compares the period-averaged concentration profiles and joint position–orientation probability density functions obtained from the two approaches for four flow oscillation frequencies, $n = 1\ \mathrm{rad\ s^{-1}}$, $n = 33.33\ \mathrm{rad\ s^{-1}}$, $n = 100\ \mathrm{rad\ s^{-1}}$ and $n = 10000\ \mathrm{rad\ s^{-1}}$, with the oscillatory flow maximum shear rate, swimming speed and rotational diffusivity held fixed at $S = 100\ \mathrm{s^{-1}}$, $V = 50\ \mu\mathrm{m\ s^{-1}}$ and $D_R = 1\ \mathrm{rad^2\ s^{-1}}$. We observe an excellent agreement between the two models over the entire frequency range considered. The 2D-FPE accurately reproduces the concentration profiles obtained from the Langevin simulations, capturing both the pronounced centerline depletion at low frequencies and the progressive recovery of a nearly uniform distribution as the oscillation frequency increases. Similarly, the probability density functions predicted by the 2D-FPE closely match those obtained from the Langevin simulations, reproducing the localization of microswimmers near the streamwise orientations $\emptyset = 0$ and $\emptyset = \pm\pi$, as well as the gradual weakening of orientational ordering with increasing oscillation frequency. We thereby establish that the 2D-FPE faithfully reproduces the underlying stochastic transport dynamics of the Langevin model.

Building on this, we analyze the variation of microswimmer concentration profiles as functions of the four non-dimensional parameters using the 2D-FPE. The observations from Langevin simulations in **Section 3.1** indicate that shear trapping in oscillatory flows is favored by slow oscillations, strong shear rates, and weak swimmer rotational diffusivities. Accordingly, the non-dimensional parameter ranges considered in the subsequent studies were selected from this regime.

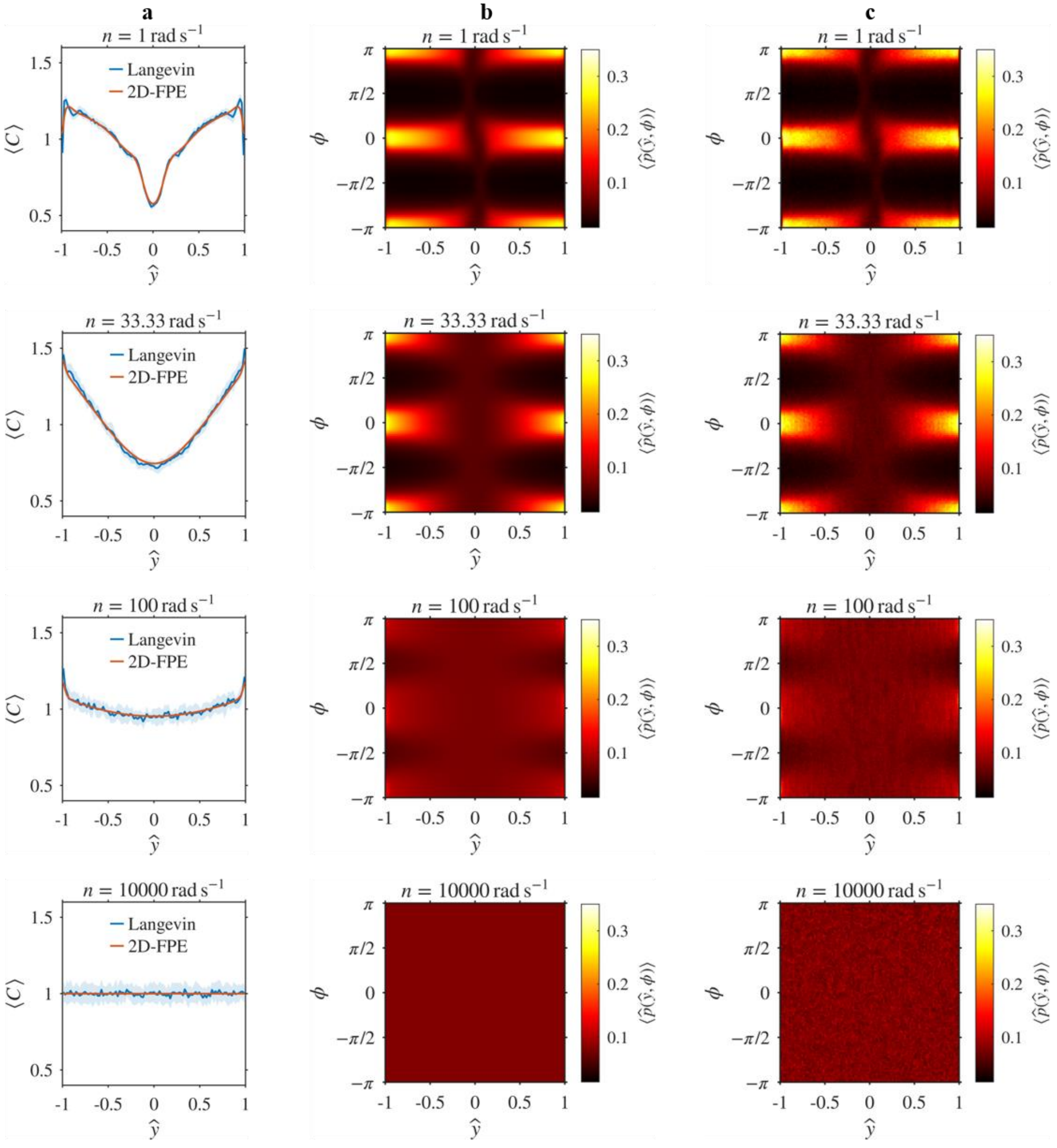

**Figure 8**. Verification of the 2D Fokker-Planck model. (a) Period-averaged concentration $\langle C \rangle$ versus $\hat{y}$, (b) period-averaged 2D-FPE joint-probability density functions $\langle \hat{p} \rangle$ versus $\emptyset$ and $\hat{y}$, and (c) period-averaged Langevin model joint-probability density functions $\langle \hat{p} \rangle$ versus $\emptyset$ and $\hat{y}$, for $S = 100\ \mathrm{s}^{-1}$, $V = 50\ \mu\mathrm{m\ s}^{-1}$, $D_R = 1\ \mathrm{rad}^2\ \mathrm{s}^{-1}$, and $n = 1, 33.33, 100,$ and $10000\ \mathrm{rad\ s}^{-1}$. Shaded regions in (a) denote one standard deviation about the mean period-averaged concentration over 20 Langevin realizations.

### 3.2.1 Suppression of shear trapping at large Womersley numbers

We here analyze the effect of the Womersley number on the microswimmer concentration distribution. **Figure 9** shows the period-averaged microswimmer concentration profiles for varying $Wo$ values between $Wo = 10^{-3}$ and $Wo = 10^{3}$, and $\beta = 0.1$, $Pe_f = 100$ and $Pe_c = 1$. For $Wo \leq 1$, we observe the concentration profiles nearly collapsing onto a single curve, exhibiting pronounced centerline depletion and wall accumulation. In this regime, recall that the flow profile is nearly parabolic and the oscillatory shear rate extends across most of the channel (c.f. **Figure 2**). This allows shear-induced alignment, and consequently preferential concentration, to develop across the channel. As $Wo$ increases further, the oscillatory shear rate becomes progressively confined to thin near-wall regions (plug-like flow profile) and leaves an increasingly weakly sheared core. As a result, only microswimmers close to the walls experience appreciable shear-induced alignment, while much of the channel remains weakly sheared and nearly isotropic. We hypothesize that the resulting reduction in the region over which shear trapping can occur progressively weakens preferential concentration, which leads to a noticeable reduction in centerline depletion at $Wo = 10$ and an almost uniform concentration profile for $Wo \geq 100$.

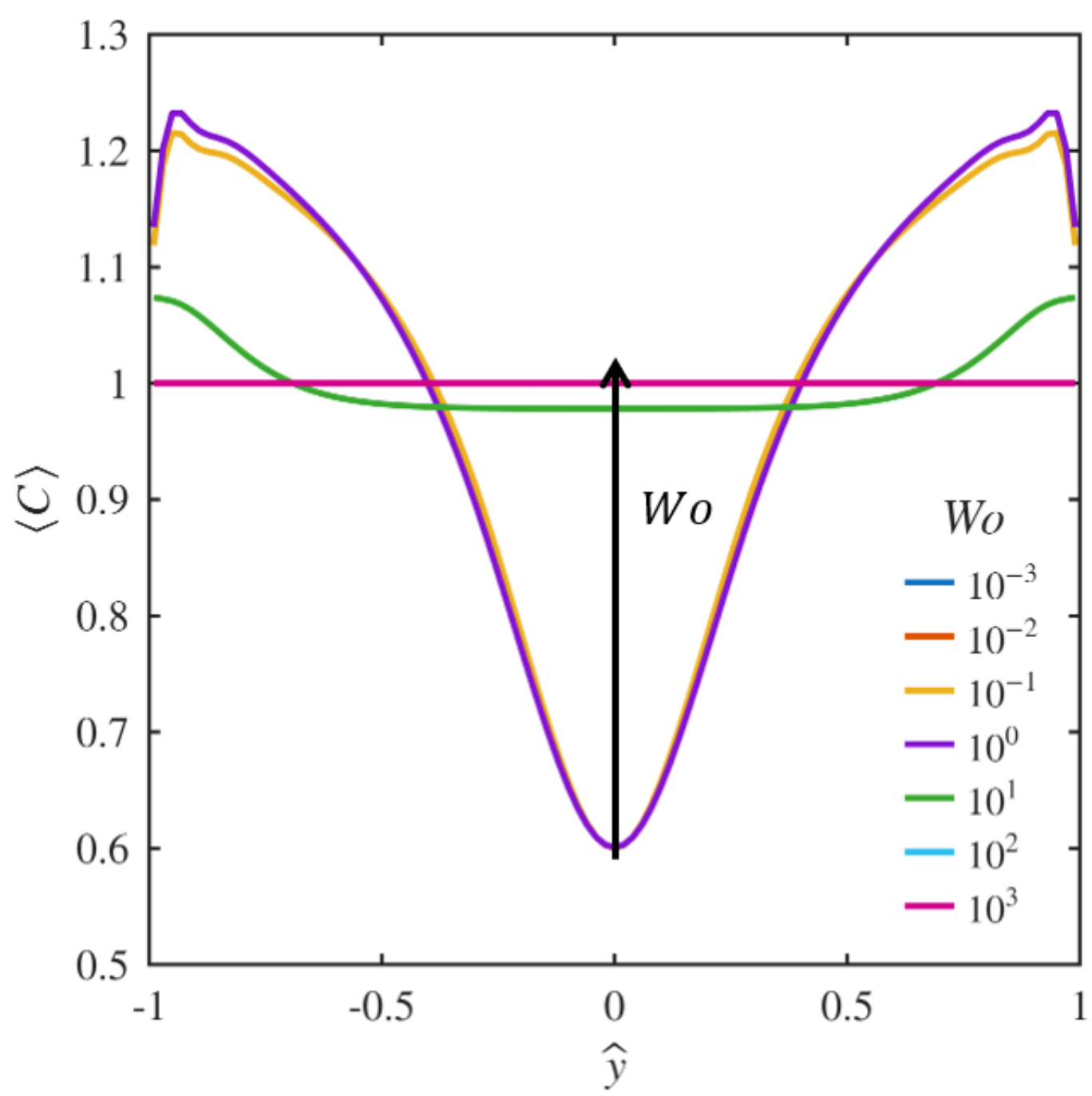

**Figure 9**. Effect of Womersley number, $Wo$, on period-averaged concentration $\langle C \rangle$ versus $\hat{y}$ for $\beta = 0.1$, $Pe_f = 100$, $Pe_c = 1$.

### 3.2.2 Competition between shear-induced alignment and rotational diffusion

We next study the effect of the flow Peclet number on the microswimmer concentration distribution. **Figure 10** shows the period-averaged microswimmer concentration profiles for varying $Pe_f$ values between $Pe_f = 1$ and $Pe_f = 10^{5}$, and $Wo = 3$, $\beta = 0.1$ and $Pe_c = 1$. For $Pe_f = 1$, we observe that the concentration profile is nearly uniform across the channel. This indicates that shear-induced reorientation is too weak relative to swimmer rotational diffusion to generate appreciable preferential concentration. As $Pe_f$ increases from $Pe_f = 1$ to $Pe_f = 10^2$ , we observe that preferential concentration becomes increasingly pronounced, with enhanced wall accumulation and broad centerline depletion. For $Pe_f > 10^2$, as $Pe_f$ increases, we observe that the wall accumulation decreases and the depletion profile undergoes a qualitative transition: the depleted region becomes progressively narrower while the centerline depletion intensifies. We characterize this large-$Pe_f$ regime by a self-similar concentration profile examined further in **Section 3.5**. Based on these observations, we infer that the extent of preferential concentration is governed by a balance between shear-induced reorientation and rotational diffusion, and a strong shear relative to diffusion modifies the microswimmer redistribution process and confines depletion to a narrow region about the channel centerline $\hat{y} = 0$.

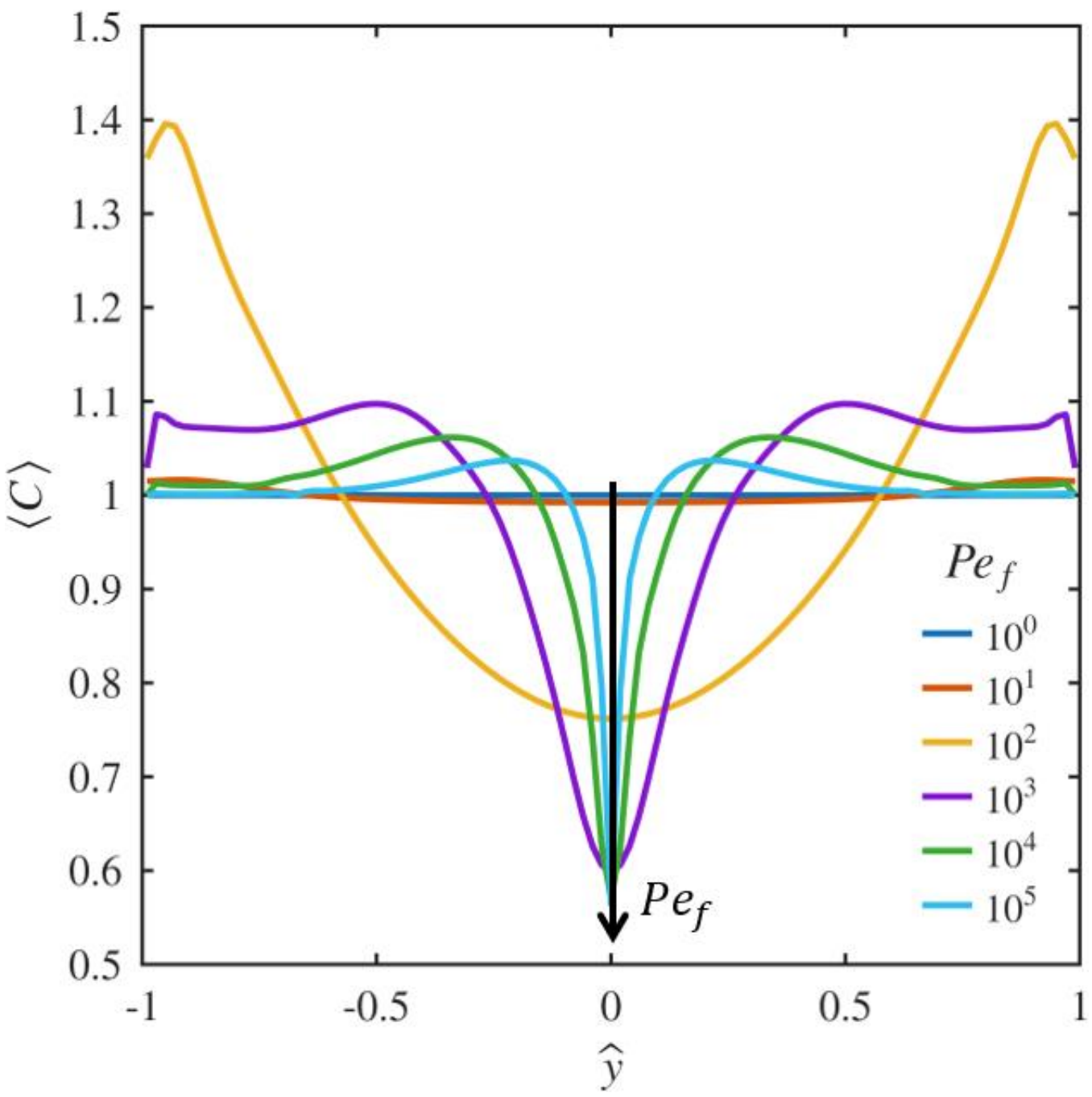

**Figure 10**. Effect of flow Peclet number, $Pe_f$, on period-averaged concentration $\langle C \rangle$ versus $\hat{y}$ for $Wo = 3$, $\beta = 0.1$, $Pe_c = 1$.

### 3.2.3 Competition between wall-normal swimming and rotational diffusion

We next examine the effect of the swim Peclet number on the microswimmer concentration distribution. **Figure 11** shows the period-averaged microswimmer concentration profiles for varying $Pe_c$ values between $Pe_c = 0$ and $Pe_c = 10^2$, and $Wo = 3$, $\beta = 0.1$ and $Pe_f = 100$. We observe that the influence of $Pe_c$ on the concentration distribution is non-monotonic. When $Pe_c = 0$ corresponding to the limit of zero swimming speed, the concentration profile is uniform. As $Pe_c$ increases from 0 to 1, we observe increasing centerline depletion and wall accumulation. We reason that as $Pe_c$ increases from 0 to 1, swimming becomes increasingly important relative to rotational diffusion, which then allows the shear-induced orientational ordering to produce stronger spatial redistribution (**Figure 6c** corresponds to $Pe_c = 0.5$). For $Pe_c > 1$, however, we observe that further increases in $Pe_c$ to $Pe_c = 10$ and $Pe_c = 10^2$ reduce preferential concentration and the concentration approaches a uniform distribution. In the large $Pe_c$ regime, we hypothesize that sufficiently large values of $Pe_c$ promote cross-stream transport that reduces the effectiveness of the oscillatory shear trapping mechanism and homogenizes the concentration field. This transition is consistent with $Pe_c\beta = V/(an)$ becoming order unity when $Pe_c = 10$, indicating that the characteristic swimming distance over one oscillation timescale becomes comparable to the channel width. Moreover, although increasing $Pe_c$ significantly changes the extent of centerline depletion and wall accumulation, we observe that the width of the depleted region remains nearly unchanged over the range considered, suggesting that the swim Peclet number primarily influences the strength rather than the spatial extent of preferential concentration.

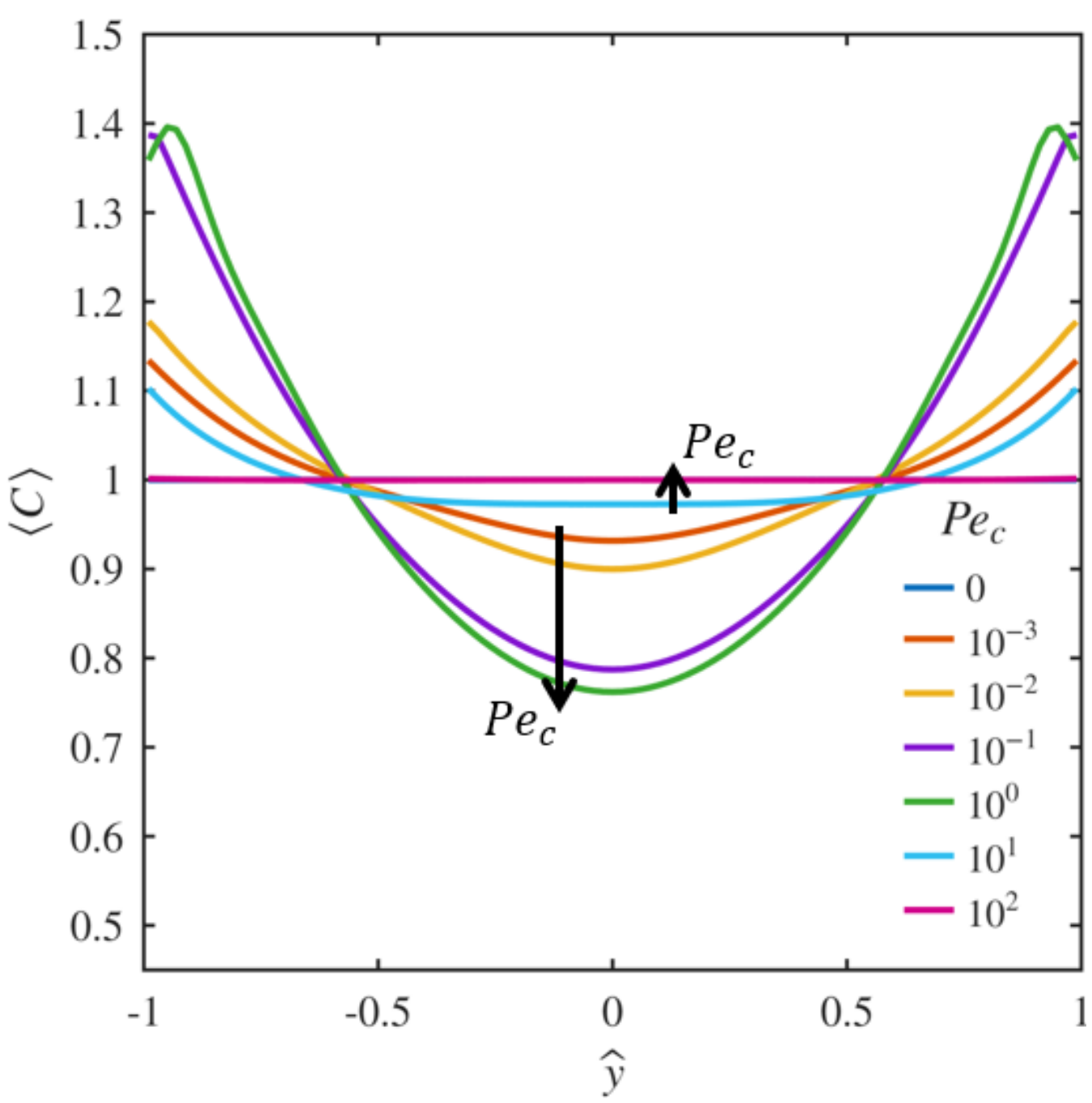

**Figure 11**. Effect of swim Peclet number, $Pe_c$, on period-averaged concentration $\langle C \rangle$ versus $\hat{y}$ for $Wo = 3$, $\beta = 0.1$, $Pe_f = 100$.

**3.2.4 Competition between rotational diffusion and frequency of oscillatory shear**

We here analyze the effect of the frequency ratio, $\beta$, on the microswimmer concentration distribution. Recall that $\beta$ is the ratio of rotational diffusion rate and flow oscillation frequency. **Figure 12** shows the period-averaged microswimmer concentration profiles for varying $\beta$ values between $\beta = 10^{-3}$ and $\beta = 10^{3}$, and $Wo = 3$, $Pe_f = 100$ and $Pe_c = 1$. We observe that preferential concentration exhibits a monotonically increasing but saturating dependence on $\beta$. As $\beta$ increases from $\beta = 10^{-3}$ to $\beta = 1$, we observe an increase in centerline depletion and wall accumulation. For $\beta > 1$, we observe that further increases in $\beta$ have nearly no effect on the concentration profile. The effect of the frequency factor $\beta$ on preferential concentration may be better understood through the Fokker-Planck equation (Eq. 2.10c). For $\beta \ll 1$, the unsteady term in the Fokker–Planck equation dominates, and temporal variations in the orientational distribution continually disrupt the build-up of the sustained anisotropy required for shear trapping. The concentration profile therefore remains nearly uniform. As $\beta$ increases for slower flow oscillations relative to rotational diffusion timescale, the influence of the unsteady term diminishes, which allows the orientational distribution to approach a quasi-steady balance between oscillatory shear-induced alignment and rotational diffusion. This leads to progressively stronger shear trapping as $\beta$ increases before the microswimmer distribution asymptotically approaches a $\beta$-independent limit. Beyond $\beta = O(1)$, we observe that the concentration profiles nearly collapse such that further increases in $\beta$ produce only marginal changes in preferential concentration. This behavior is consistent with the asymptotic solution derived in **Section 3.5**, which predicts that the depletion index approaches a $\beta$-independent limit for $\beta \gg 1$ (**Section 3.5**).

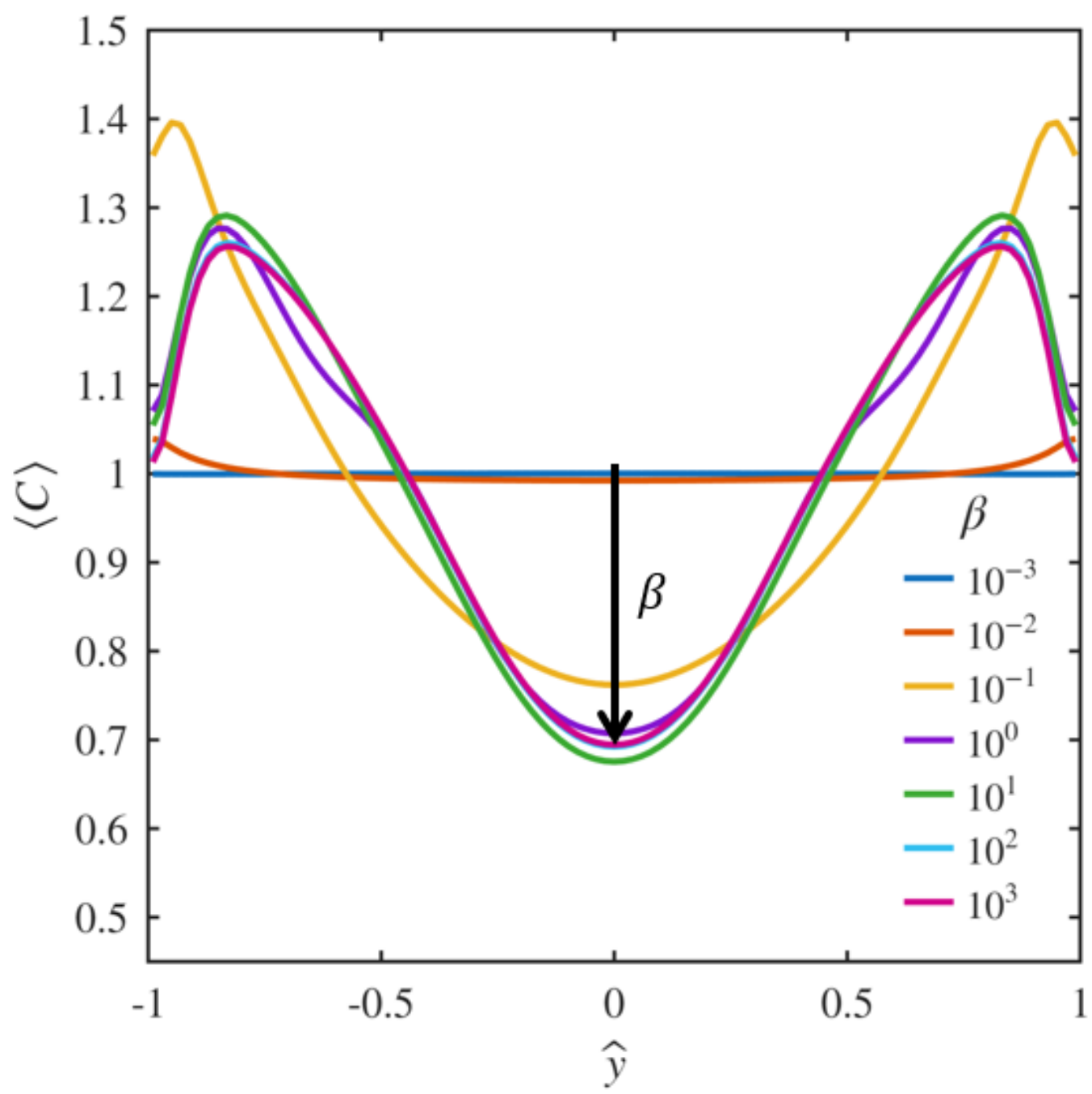

**Figure 12**. Effect of the frequency ratio, $\beta$, on period-averaged concentration $\langle C \rangle$ versus $\hat{y}$ for $Wo = 3$, $Pe_f = 100$, $Pe_c = 1$.

### 3.3 One-dimensional Fokker-Planck equation

In **Section 2.4**, we derived the one-dimensional Fokker–Planck equation under the assumption of weak swimming ($Pe_c \ll 1$) and it describes only the orientational dynamics of the microswimmer population at a given $\hat{y}$-location. We verify the reduced model by comparing the period-averaged orientational probability density function, $\langle \hat{p}(\emptyset; \hat{y}) \rangle$, predicted by the 1D-FPE with that obtained from the Langevin simulations at two representative oscillation frequencies, $n = 1\ \mathrm{rad\ s^{-1}}$ and $n = 100\ \mathrm{rad\ s^{-1}}$, and a zero swimming speed ($Pe_c = 0$). In **Figure 13**, we observe excellent agreement between the two approaches at both the low and high frequency limits. The 1D-FPE results accurately reproduce the strong localization of microswimmers near the streamwise orientations $\emptyset = 0$ and $\emptyset = \pm\pi$ at $n = 1\ \mathrm{rad\ s^{-1}}$, as well as the weaker orientational ordering at $n = 100\ \mathrm{rad\ s^{-1}}$. The close correspondence between the 1D-FPE and Langevin predictions confirms that the reduced formulation provides an accurate description of the orientational distribution when swimming is weak.

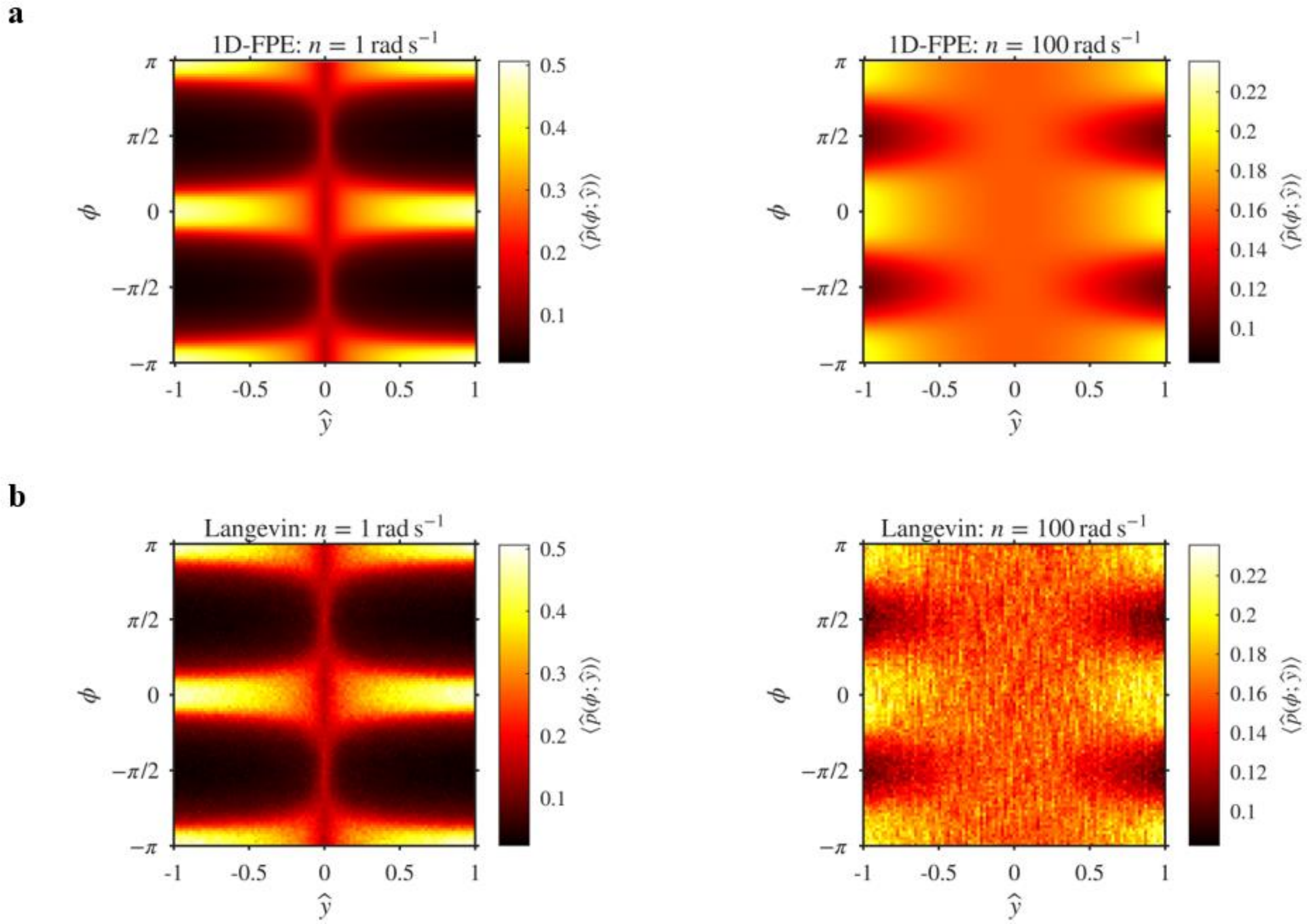

**Figure 13**. Verification of the one-dimensional Fokker-Planck model: period-averaged orientational probability density function $\langle\hat{p}(\emptyset;\hat{y})\rangle$ versus $\emptyset$ and $\hat{y}$, for $n = 1\ \mathrm{rad\ s^{-1}}$ and $n = 100\ \mathrm{rad\ s^{-1}}$, for $S = 100\ \mathrm{s^{-1}}$, $V = 0\ \mu\mathrm{m\ s^{-1}}$ and $D_R = 1\ \mathrm{rad^2\ s^{-1}}$, obtained from (a) 1D-FPE, and (b) Langevin model.

Having verified the reduced model, we now examine its range of validity by comparing the 1D-FPE results of $\langle\hat{p}(\emptyset;\hat{y})\rangle$ with Langevin simulations at increasing swimming speeds. **Figure 14** shows the orientational probability density functions obtained from the 1D-FPE and Langevin simulations for $V = 0\ \mu\mathrm{m\ s^{-1}}$, $V = 5\ \mu\mathrm{m\ s^{-1}}$, $V = 50\ \mu\mathrm{m\ s^{-1}}$ and $V = 500\ \mu\mathrm{m\ s^{-1}}$ ($Pe_c = 0$, $Pe_c = 0.05$, $Pe_c = 0.5$ and $Pe_c = 5$), with the other parameters held fixed at $n = 1\ \mathrm{rad\ s^{-1}}$, $S = 100\ \mathrm{s^{-1}}$ and $D_R = 1\ \mathrm{rad^2\ s^{-1}}$. For $V = 0\ \mu\mathrm{m\ s^{-1}}$ and $V = 5\ \mu\mathrm{m\ s^{-1}}$, we observe the distributions from Langevin simulations closely match the 1D-FPE result. Noticeable deviations begin to emerge at $V = 50\ \mu\mathrm{m\ s^{-1}}$, while at $V = 500\ \mu\mathrm{m\ s^{-1}}$, the Langevin distribution exhibits appreciable probability at intermediate orientations and a markedly different spatial structure, not captured by the 1D-FPE. These deviations arise because the assumptions underlying the 1D-FPE become progressively less accurate as the swimming speed increases. We hypothesize that strong swimming couples the orientational and positional dynamics, which is neglected by the reduced model. Consequently, the 1D-FPE provides an accurate approximation only in the weak-swimming regime, corresponding to $Pe_c \ll 1$.

**a 1D-FPE**

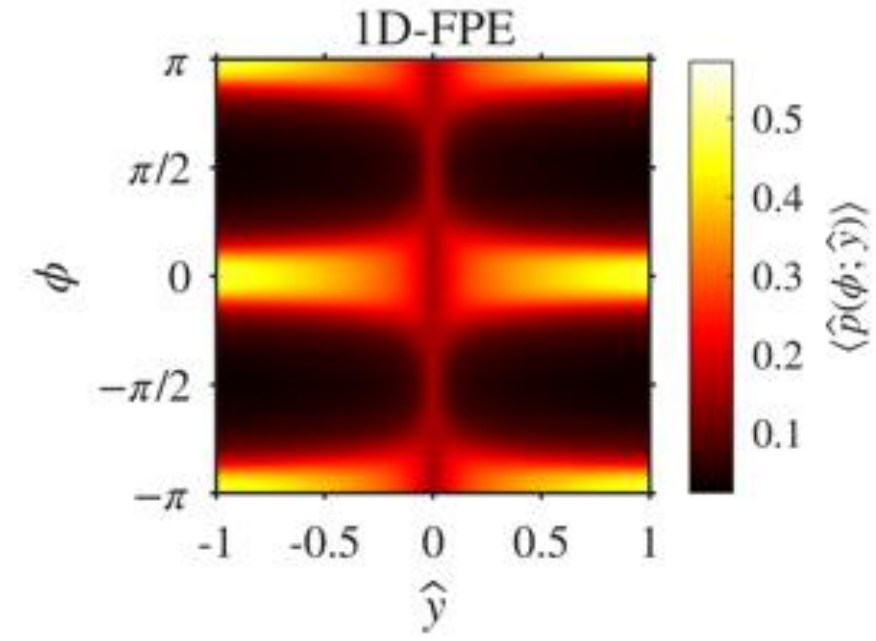

**b Langevin**

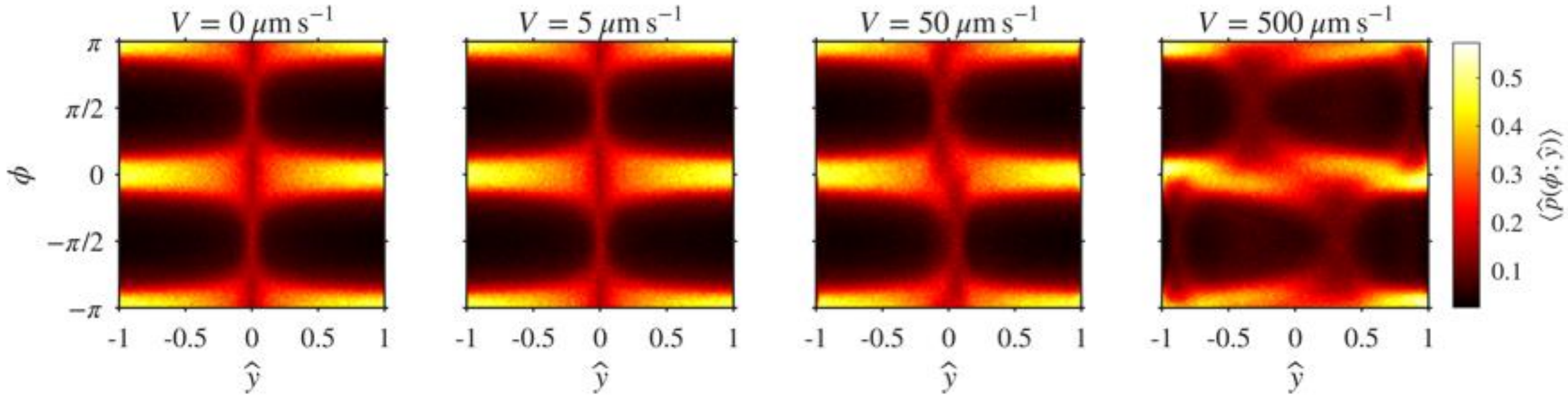

**Figure 14**. Validity of the one-dimensional Fokker-Planck model: period-averaged orientational probability density function $\langle\hat{p}(\emptyset;\hat{y})\rangle$ versus $\emptyset$ and $\hat{y}$, for $V = 0\ \mu\mathrm{m\ s^{-1}}$, $V = 5\ \mu\mathrm{m\ s^{-1}}$, $V = 50\ \mu\mathrm{m\ s^{-1}}$ and $V = 500\ \mu\mathrm{m\ s^{-1}}$, for $n = 1\ \mathrm{rad\ s^{-1}}$, $S = 100\ \mathrm{s^{-1}}$ and $D_R = 1\ \mathrm{rad^2\ s^{-1}}$, obtained from (a) 1D-FPE, and (b) Langevin model.

We employ the 1D-FPE in the following sections to derive analytical approximations for the normalized period-averaged concentration profile and depletion index.

### 3.4. Harmonic response of the orientational distribution

We here analytically characterize the local orientational response of microswimmers to oscillatory shear by considering the one-dimensional Fokker–Planck equation (Eq. 2.14b) at a fixed cross-stream position $\hat{y}$. The orientational probability density $\hat{p}(\emptyset,\tau;\hat{y})$ satisfies,

$$\frac{\partial\hat{p}}{\partial\tau} = -\frac{\partial}{\partial\emptyset}[K(\tau)(\alpha\cos 2\emptyset - 1)\hat{p}] + \beta\frac{\partial^2\hat{p}}{\partial\emptyset^2}, \tag{3.1}$$

where, $K(\tau) = \frac{1}{2}Pe_f\beta\,\Sigma(\tau;\hat{y},Wo)$, such that $\Sigma(\tau;\hat{y},Wo) = \frac{1}{Wo}\{F_1(\hat{y};Wo)\cos\tau + F_2(\hat{y};Wo)\sin\tau\}$. At a specified $\hat{y}$, without loss of generality, the local time origin may be shifted such that the oscillatory shear is written as $\Sigma(\tau;\hat{y},Wo) = \Sigma_a(\hat{y},Wo)\cos\tau$, where $\Sigma_a = \mathrm{Amp}[\Sigma]$ is the local shear amplitude.

Given the periodic nature of the problem, we may describe the orientational distribution through its Fourier moments, $C_n(\tau) = \langle\cos n\emptyset\rangle$, $S_n(\tau) = \langle\sin n\emptyset\rangle$. The corresponding orientational probability density may be reconstructed from the Fourier expansion,

$$\hat{p}(\emptyset,\tau) = \frac{1}{2\pi}\left[1+2\ \sum_{n=1}^{\infty}(C_n(\tau)\cos n\emptyset + S_n(\tau)\sin n\emptyset)\right]. \tag{3.2}$$

For any smooth $2\pi$-periodic function $f(\emptyset)$, multiplication of Eq. 3.1 by $f$, followed by integration over $\emptyset$, gives,

$$\frac{d\langle f\rangle}{d\tau} = \langle K(\tau)(\alpha\cos 2\emptyset - 1)f'(\emptyset) + \beta f''(\emptyset)\rangle, \tag{3.3}$$

where, $\langle f\rangle = \int_{-\pi}^{\pi} f(\emptyset)\,\hat{p}(\emptyset,\tau)\,d\emptyset$ . Taking $f = \cos n\emptyset$ and $f = \sin n\emptyset$ yields the following harmonic hierarchy,

$$\begin{aligned} \dot{C}_n &= nKS_n - \frac{n\alpha K}{2}(S_{n+2}+S_{n-2}) - n^2\beta C_n\,, \\ \dot{S}_n &= \frac{n\alpha K}{2}(C_{n+2}+C_{n-2}) - nKC_n - n^2\beta S_n\,. \end{aligned} \tag{3.4}$$

Thus, the oscillatory shear couples neighboring even orientational harmonics. In particular, the $n$-th sine or cosine mode is coupled to the modes, $n-2$ and $n+2$.

For $n = 2$, using $C_0 = 1$ and $S_0 = 0$, we get,

$$\begin{aligned} \dot{C}_2 + 4\beta C_2 &= 2KS_2 - \alpha KS_4\,, \\ \dot{S}_2 + 4\beta S_2 &= \alpha K + \alpha KC_4 - 2KC_2\,. \end{aligned} \tag{3.5a}$$

The corresponding equations for the fourth harmonics are,

$$\begin{aligned} \dot{C}_4 + 16\beta C_4 &= 4KS_4 - 2\alpha K(S_6 + S_2)\,, \\ \dot{S}_4 + 16\beta S_4 &= 2\alpha K(C_6 + C_2) - 4KC_4\,. \end{aligned} \tag{3.5b}$$

These equations reveal the sequence in which the orientational moments are generated. For an initially isotropic distribution, $C_n = S_n = 0$ for all $n \geq 1$, while $C_0 = 1$. Consequently, $S_2$ is the only moment directly forced by the oscillatory shear through the term $\alpha K$, and therefore constitutes the leading orientational response. In contrast, the equation for $C_2$ has no direct forcing from the isotropic state but is generated through the product $KS_2$. Similarly, $C_4$ is first generated by the coupling $KS_2$, while $S_4$ requires an additional multiplication by $K$. Therefore, the hierarchy implies $S_2 = O(K)$, $C_2 = O(K^2)$, $C_4 = O(K^2)$ and $S_4 = O(K^3)$.

To obtain an analytical approximation of the harmonic response, we consider the weak-forcing limit ($Pe_f \ll 1$) and introduce a regular perturbation expansion in $Pe_f$, while holding $\beta$, $\alpha$, and $\Sigma_a$ fixed. Since $K(\tau) = \frac{1}{2}Pe_f\beta\Sigma_a\cos\tau$, the moment expansions consistent with the hierarchy are given by,

$$S_2 = Pe_f s_1 + Pe_f^3 s_3 + O\left(Pe_f^5\right),$$
$$C_2 = Pe_f^2 c_2 + O\left(Pe_f^4\right),$$
$$C_4 = Pe_f^2 c_4 + O\left(Pe_f^4\right), \tag{3.6}$$
$$S_4 = Pe_f^3 s_4 + O\left(Pe_f^5\right).$$

Here, uppercase symbols denote the physical orientational moments, whereas the corresponding lowercase symbols represent the $O(1)$ coefficient functions at each order in the perturbation expansion.

The equations for $s_1$ at $O(Pe_f)$, $c_2$ and $c_4$ at $O(Pe_f^2)$, and $s_3$ at $O(Pe_f^3)$, respectively are given by,

$$\dot{s}_1 + 4\beta s_1 = \alpha\gamma \cos\tau\,,$$
$$\dot{c}_2 + 4\beta c_2 = 2\gamma \cos\tau\, s_1\,,$$
$$\dot{c}_4 + 16\beta c_4 = -2\alpha\gamma \cos\tau\, s_1\,, \tag{3.7}$$
$$\dot{s}_3 + 4\beta s_3 = \alpha\gamma \cos\tau\, c_4 - 2\gamma \cos\tau\, c_2\,,$$

where, $\gamma = \frac{1}{2}\beta\Sigma_a$. Eqs. 3.7 are solved to obtain the following long-time periodic solutions (for $s_3$, only the first temporal harmonic is presented):

$$s_1(\tau) = \frac{\alpha\gamma}{B}\left[4\beta\cos\tau + \sin\tau\right],$$
$$c_2(\tau) = \frac{\alpha\gamma^2}{B}\left[1 + \frac{8\beta^2 - 1}{2A}\cos 2\tau + \frac{3\beta}{A}\sin 2\tau\right],$$
$$c_4(\tau) = \frac{\alpha^2\gamma^2}{B}\left[-\frac{1}{4} + \frac{1 - 32\beta^2}{2C}\cos 2\tau - \frac{6\beta}{C}\sin 2\tau\right], \tag{3.8}$$
$$s_3(\tau) = -\frac{3\alpha\gamma^3}{2AB}\left[2\beta\left\{1 + \frac{\alpha^2 A(32\beta^2 - 1)}{BC}\right\}\cos\tau + \left\{1 + \frac{24\alpha^2\beta^2 A}{BC}\right\}\sin\tau\right],$$

where, $A = 1 + 4\beta^2$, $B = 1 + 16\beta^2$, and $C = 1 + 64\beta^2$.

Therefore, the first temporal harmonic of $S_2$ is given by,

$$S_2(\tau) = Pe_f[U_1\cos\tau + V_1\sin\tau] + Pe_f^3[U_3\cos\tau + V_3\sin\tau] + O\left(Pe_f^5\right) \tag{3.9}$$

where, $U_1 = \frac{\alpha\gamma}{B}4\beta$, $V_1 = \frac{\alpha\gamma}{B}$, $U_3 = -\frac{3\alpha\gamma^3}{2AB}2\beta\left\{1 + \frac{\alpha^2 A(32\beta^2-1)}{BC}\right\}$, and $V_3 = -\frac{3\alpha\gamma^3}{2AB}\left\{1 + \frac{24\alpha^2\beta^2 A}{BC}\right\}$.

Since $S_2(\tau)$ is the first angular harmonic directly forced by the oscillatory shear, its first temporal harmonic constitutes the fundamental orientational response of the microswimmers. Accordingly, the orientational response to oscillatory shear may be recast in amplitude-phase form as follows:

$$S_2(\tau) = A_{S_2} \cos\left(\tau - \delta_{S_2}\right) + O\left(Pe_f^5\right), \tag{3.10a}$$

where,

$$\frac{A_{S_2}}{\frac{1}{2} Pe_f \beta \Sigma_a} = \frac{\alpha}{\sqrt{1+16\beta^2}} \left[1 - Pe_f^2 \frac{3\gamma^2[16\beta^2\alpha^2 A + C(1+8\beta^2)]}{2ABC}\right] + O\left(Pe_f^4\right),$$
$$\delta_{S_2} = \tan^{-1}\left(\frac{1}{4\beta}\right) - Pe_f^2 \frac{3\beta\gamma^2(\alpha^2 A + C)}{ABC} + O\left(Pe_f^4\right). \tag{3.10b}$$

Here, the amplitude is obtained directly from the Euclidean norm of the cosine and sine coefficients, while the phase lag follows from an asymptotic expansion of $\tan^{-1}[(Pe_f V_1 + Pe_f^3 V_3)/(Pe_f U_1 + Pe_f^3 U_3)]$. At leading order, the orientational response normalized by the forcing amplitude, $\frac{1}{2} Pe_f \beta \Sigma_a$, can be described by the linear transfer law,

$$\hat{S}_2(\tau) = \frac{\alpha}{\sqrt{1+16\beta^2}} \cos\left[\tau - \tan^{-1}\left(\frac{1}{4\beta}\right)\right], \tag{3.10c}$$

whose gain and phase lag, for a given microswimmer shape, depend only on the dimensionless rotational response time $\beta$. Specifically, the gain and phase lag are $\alpha/\sqrt{1+16\beta^2}$ and $\tan^{-1}(1/4\beta)$ respectively. Consequently, all weakly forced suspensions collapse onto a universal normalized response described in Eq. 3.10c, with finite forcing introducing $O(Pe_f^2)$ nonlinear corrections to both the gain and the phase lag. Since these corrections are always negative, the finite-$Pe_f$ gain and phase lag are both reduced relative to their leading-order predictions and approach the linear-response limits asymptotically as $Pe_f \to 0$.

### 3.5. Depletion Index

We next consider an integral measure of the period-averaged concentration distribution to quantify the long-time preferential concentration of microswimmers in oscillatory flow and facilitate comparison between the 2D Fokker–Planck and asymptotic results. In particular, we introduce a depletion index that provides a scalar measure of the extent of microswimmer depletion from the central region of the channel. To this end, we substitute the perturbation solutions for the Fourier moments (Eqs. 3.6 and 3.8) into the Fourier representation of $\hat{p}$ (Eq. 3.2), which yields the following asymptotic solution of the one-dimensional Fokker–Planck equation (Eq. 2.14b), valid up to $O(Pe_f^2)$:

$$\hat{p}(\emptyset,\tau;\hat{y}) = \frac{1}{2\pi}\left\{1 + \frac{2Pe_f\alpha\gamma}{B}(4\beta\cos\tau + \sin\tau)\sin 2\emptyset \right.$$
$$+ \frac{2Pe_f^2\alpha\gamma^2}{B}\left[1 + \frac{8\beta^2 - 1}{2A}\cos 2\tau + \frac{3\beta}{A}\sin 2\tau\right]\cos 2\emptyset \qquad (3.11)$$
$$\left. + \frac{2Pe_f^2\alpha^2\gamma^2}{B}\left[-\frac{1}{4} + \frac{1 - 32\beta^2}{2C}\cos 2\tau - \frac{6\beta}{C}\sin 2\tau\right]\cos 4\emptyset\right\} + O\left(Pe_f^3\right).$$

This analytical expression forms the basis for deriving the depletion index in the weak-forcing regime. In the limit of low $Wo$, the local oscillatory shear amplitude may be written as $\Sigma_a(\hat{y}, Wo) \cong -\hat{y}\left[1 + \frac{Wo^4}{6}(3 - \hat{y}^2)^2\right]^{1/2} \cong -\hat{y}$ (c.f. Eqs. 3.1, 2.10b).

The 1D-FPE is invariant under the transformation $\emptyset \to \emptyset + \pi$. Since the isotropic initial condition, $\hat{p}(\emptyset, 0) = 1/(2\pi)$, is also invariant under this transformation, the solution satisfies $\hat{p}(\emptyset + \pi, \tau) = \hat{p}(\emptyset, \tau)$, implying that the orientational probability density is $\pi$-periodic. However, in the 2D-FPE, this symmetry is broken by the wall-normal swimming term, $Pe_c \sin\emptyset\, \partial\hat{p}/\partial\hat{y}$, which is not invariant under the transformation $\emptyset \to \emptyset + \pi$. Note that the $\pi$-periodicity of $\hat{p}(\emptyset, \tau; \hat{y})$ in the 1D-FPE is apparent in **Figure 13**. Following this, the effective period-averaged wall-normal swimming speed of microswimmers across the channel is given by $V_{\mathrm{EFF}}(\hat{y}) = \frac{1}{2\pi}\int_0^{2\pi}\int_0^{\pi} V\sin\emptyset\, \hat{p}(\emptyset, \tau; \hat{y})\, d\emptyset\, d\tau$, and the period-averaged microswimmer density, $B'(\hat{y})$, is given by $B'(\hat{y})/B'(0) = V_{\mathrm{EFF}}(0)/V_{\mathrm{EFF}}(\hat{y})$ (Rusconi et al., 2014; Schnitzer, 1993). Therefore, we obtain,

$$\frac{B'(\hat{y})}{B'(0)} = \frac{V_{eff}(0)}{V_{eff}(\hat{y})} = \left\{1 - \sigma\, Pe_f^2 \frac{\beta^2}{(1 + 16\beta^2)}\hat{y}^2\right\}^{-1} \cong 1 + \sigma\, Pe_f^2 \frac{\beta^2}{(1 + 16\beta^2)}\hat{y}^2, \qquad (3.12)$$

where $\sigma = \frac{\alpha}{6}\left\{1 - \frac{\alpha}{20}\right\}$, and the approximation, $(1 - x)^{-1} \cong 1 + x;\ x \ll 1$, has been used since $Pe_f \ll 1$. The microswimmer density is normalized by the mean microswimmer density across the channel to obtain the analytical expression for the (period-averaged) normalized microswimmer concentration.

$$\langle C(\hat{y})\rangle = \frac{B'(\hat{y})}{\frac{1}{2}\int_{-1}^{1} B'(\hat{y})\, \mathrm{d}\hat{y}} = \frac{1 + \sigma\, Pe_f^2 \frac{\beta^2}{(1 + 16\beta^2)}\hat{y}^2}{1 + \frac{1}{3}\sigma\, Pe_f^2 \frac{\beta^2}{(1 + 16\beta^2)}} \qquad (3.13)$$

We define the depletion index, $I_D$, as the integrated deficit of the normalized period-averaged concentration below its mean value.

$$I_D = \int_{-\hat{y}^*}^{\hat{y}^*} [1 - \langle C(\hat{y})\rangle]\, \mathrm{d}\hat{y} = 2\int_0^{\hat{y}^*} [1 - \langle C(\hat{y})\rangle]\, \mathrm{d}\hat{y}, \qquad (3.14)$$

where the halfwidth of depletion zone, $\hat{y}^*$, corresponds to $\langle C(\hat{y}^*)\rangle = 1$. Setting $\langle C(\hat{y})\rangle = 1$ in Eq. 3.13, we obtain $\hat{y}^* = 1/\sqrt{3} + O(Wo^4)$. This suggests that the width of the depletion zone is practically constant when $Pe_f \ll 1$, $Wo \ll 1$ and $Pe_c \ll 1$. We evaluate the integral in Eq. 3.14 to obtain the following analytical expression for microswimmer depletion index for weak oscillatory flows:

$$I_D \cong \underbrace{\frac{\sigma}{64} Pe_f^2}_{\text{steady}} \underbrace{\frac{16\beta^2}{1 + 16\beta^2}}_{\text{oscillatory}} \tag{3.15}$$

Notably, the quadratic dependence of $I_D$ on $Pe_f$ follows the weak-shear scaling reported previously by Rusconi et al. (2014) for steady flows, while the effect of suppression of depletion due to flow oscillations is captured by the factor $16\beta^2/(1 + 16\beta^2)$. In particular, the perturbation solution predicts that the $Pe_f^2$ scaling is suppressed by effectively a factor $16\beta^2$ in the rapid-oscillation limit ($\beta \ll 1$), which captures a frequency-dependent attenuation regime of shear-induced alignment. Furthermore, in the limit $\beta \gg 1$, the oscillatory-factor approaches unity, where the attenuation response becomes insensitive to changes in oscillation frequency.

**Figures 15** shows period-averaged normalized microswimmer concentration profiles obtained using perturbation analysis (Eq. 3.13) compared against the 2D-FPE numerical results for $Pe_f = 0.1$, $Pe_f = 1$ and $Pe_f = 5$, and $Wo = 0.1$, $Pe_c = 0.01$ and $\beta = 1$. We observe a good comparison of the perturbation solution with the 2D-FPE predictions in the weak-shear regime, $Pe_f \lesssim 1$.

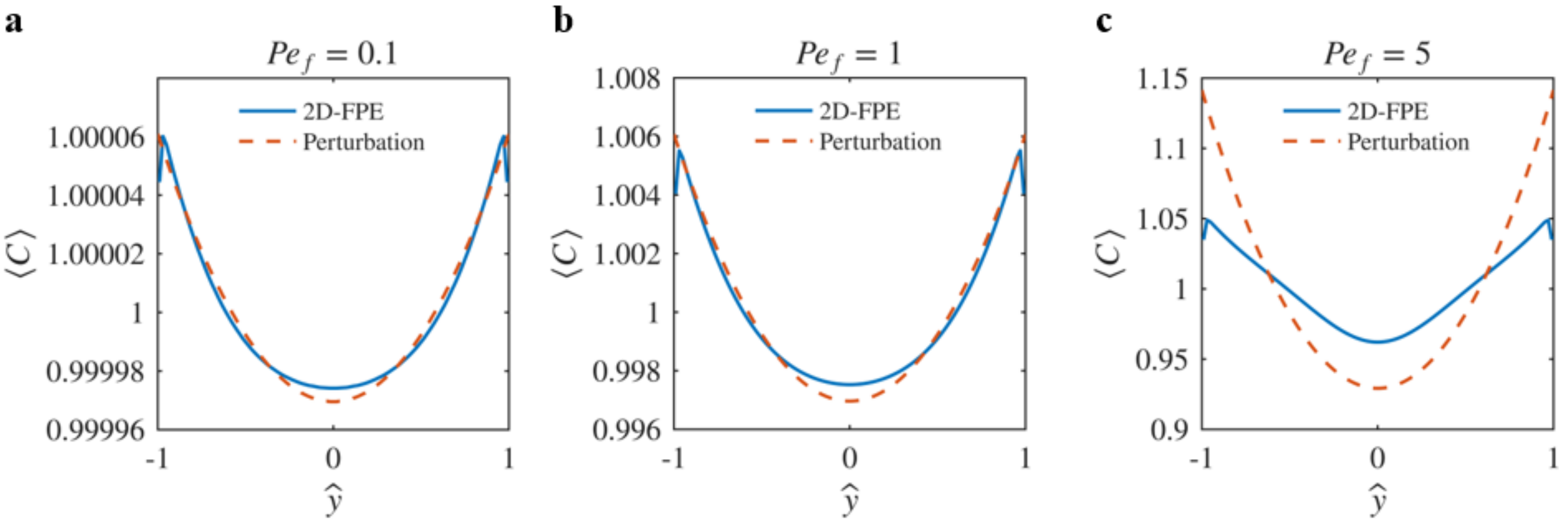

**Figure 15**. Verification of the perturbation concentration profile: period-averaged concentration $\langle C\rangle$ versus $\hat{y}$ obtained using perturbation analysis and the 2D-FPE model for (a) $Pe_f = 0.1$, (b) $Pe_f = 1$, and (c) $Pe_f = 5$, for $Wo = 0.1$, $Pe_c = 0.01$ and $\beta = 1$.

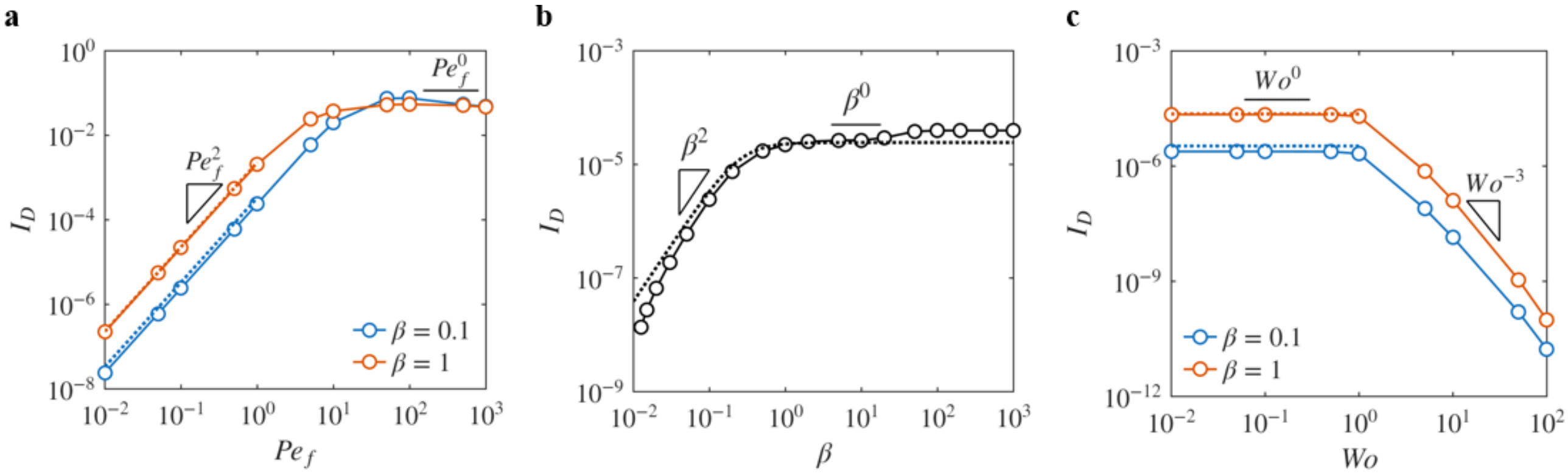

**Figure 16**. Verification of the perturbation depletion index and asymptotic scaling laws: depletion index $I_D$ versus (a) $Pe_f$, for $\beta = 0.1$ and $\beta = 1$, for $Pe_c = 0.01$ and $Wo = 0.1$, (b) $\beta$, for $Pe_c = 0.01$, $Wo = 0.1$ and $Pe_f = 0.1$, and (c) $Wo$, for $\beta = 0.1$ and $\beta = 1$, for $Pe_c = 0.01$ and $Pe_f = 0.1$. Symbols connected by solid lines denote the numerical 2D-FPE results, and dotted lines represent the perturbation solution.

**Figure 16** shows depletion index obtained from 2D-FPE and perturbation solutions versus the governing dimensionless parameters, $Pe_f$, $\beta$ and $Wo$. In the weak-shear limit ($Pe_f \ll 1$), **Figure 16a** shows the quadratic dependence $I_D \sim Pe_f^2$, consistent with the perturbation solution (Eq. 3.15). A decrease in the frequency ratio from $\beta = 1$ to $\beta = 0.1$, which corresponds to more rapid flow oscillations relative to rotational diffusion, shifts the quadratic regime to lower magnitudes of depletion index but preserves the $Pe_f^2$ scaling. Note that this trend was predicted by results from perturbation analysis (Eq. 3.15), and we here in **Figure 16** show its validity by comparing the asymptotic limit with the full 2D-FPE solution.

For $Pe_f \gtrsim 1$, where the perturbation approximation ceases to apply, we see that the depletion index approaches an approximately $Pe_f$-independent plateau as $Pe_f$ increases. We examine the origin of this behavior by observing a self-similar structure of the period-averaged microswimmer concentration profiles in the large-$Pe_f$ regime. **Figure 17** shows the period-averaged concentration versus $\hat{y}$ for $50 \leq Pe_f \leq 1000$ and $\beta = 1$, and the corresponding rescaled profiles plotted in terms of the similarity variables $C^* = 1 - (1 - \langle C \rangle)/A$ and $\eta = \hat{y}/\hat{y}^*$, where $A = 1 - \langle C(\hat{y} = 0) \rangle$ denotes the period-averaged centerline depletion and $\hat{y}^*$ denotes the depletion halfwidth corresponding to $\langle C(\hat{y}^*) \rangle = 1$. From **Figure 17a**, we observe that the centerline depletion becomes progressively deeper and the depletion zone is simultaneously more confined around the channel centerline as $Pe_f$ increases. We derive the following empirical fit for the period-averaged microswimmer concentration profiles in the depletion region in the large-$Pe_f$ regime:

$$\langle C \rangle = 1 - A \left[ 1 - \left( \frac{|\hat{y}|}{\hat{y}^*} \right)^{0.8} \right]^{1.5} \; ; \; |\hat{y}| \leq \hat{y}^* \tag{3.16}$$

where the centerline depletion $A$ and depletion halfwidth $\hat{y}^*$ are observed to scale as $A = 0.13Pe_f^{0.18}$ and $\hat{y}^* = 0.52Pe_f^{-0.17}$, respectively. When rescaled by $A$ and $\hat{y}^*$, the concentration profiles $C^*$ versus $\eta$ collapse onto a single master curve for $50 \leq Pe_f \leq 1000$ (**Figure 17b**).

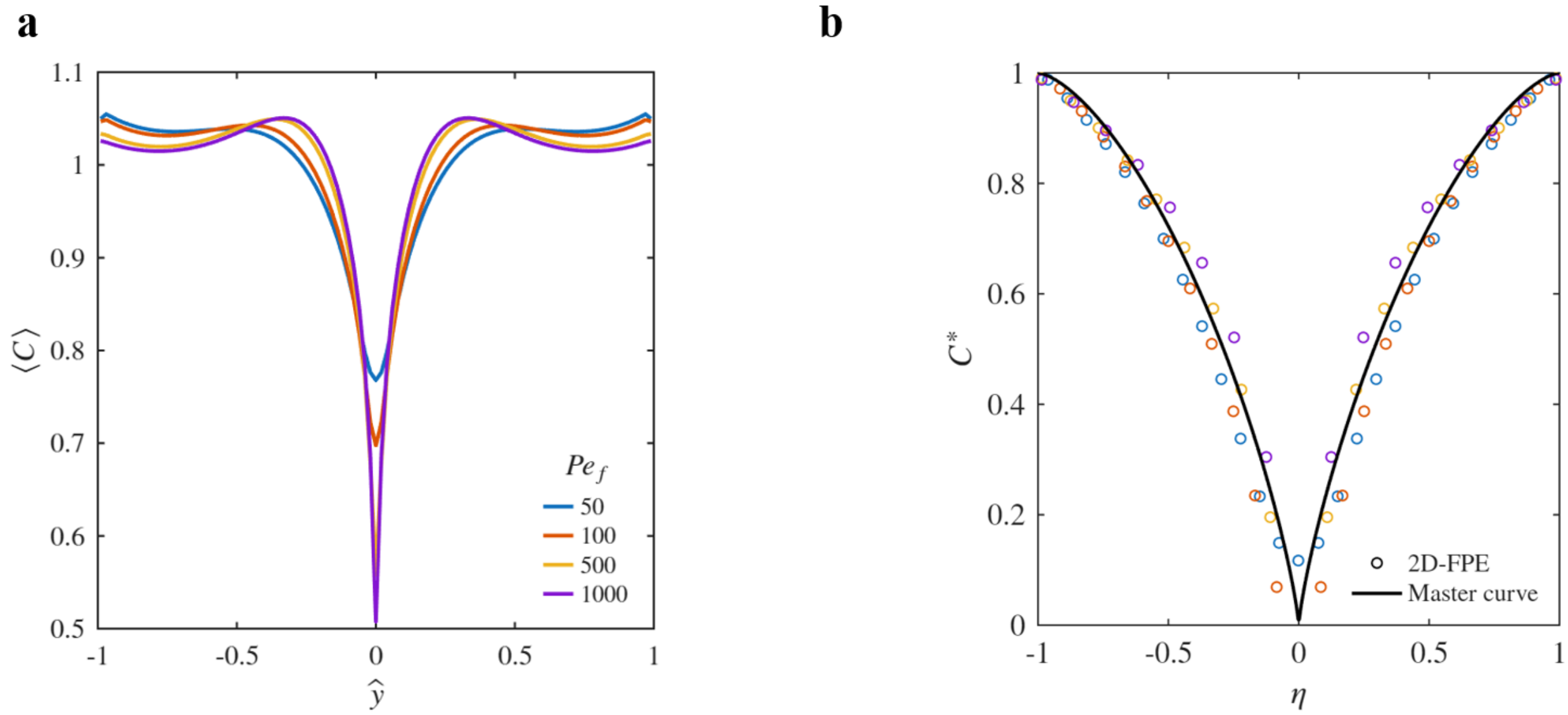

**Figure 17**. Self-similar concentration profiles in the large-$Pe_f$ regime: (a) effect of flow Peclet number on period-averaged concentration $\langle C \rangle$ versus $\hat{y}$, and (b) collapse of the rescaled concentration profiles $C^*$ versus $\eta$ under the scaling $C^* = 1 - (1 - \langle C \rangle)/A$ and $\eta = \hat{y}/\hat{y}^*$, for $\beta = 1$, $Pe_c = 0.01$ and $Wo = 0.1$.

Furthermore, since depletion index is given by the integrated concentration deficit (Eq. 3.14), we obtain $I_D = 2\int_0^{\hat{y}^*}(1 - \langle C \rangle)\,d\hat{y} \sim A\hat{y}^* \sim Pe_f^{0.01}$, explaining the observed plateau of $I_D$ at large $Pe_f$. The increase in centerline depletion is compensated by a nearly equal narrowing of the depletion width, which leaves the depletion index asymptotically independent of $Pe_f$ in the large-$Pe_f$ regime.

**Figure 16b** shows the dependence of the depletion index on the frequency ratio $\beta$. Recall that $\beta$ is the ratio of rotational diffusivity $D_R$ and flow oscillation frequency $n$. The perturbation solution shows good agreement with the 2D-FPE results over the entire range of $\beta$ considered. For $\beta \ll 1$, we recover the scaling $I_D \sim \beta^2$ which highlights the suppression of microswimmer depletion in the regime of rapid flow oscillations relative to the microswimmer rotational diffusion timescale. As $\beta$ increases beyond order unity, we see that $I_D$ approaches a $\beta$-independent plateau, $I_D \sim \beta^0$, as predicted by the asymptotic solution (Eq. 3.15).

Finally, **Figure 16c** shows that depletion index is independent of the Womersley number at small $Wo$ as predicted by the perturbation solution, whereas for $Wo > O(1)$, it decreases sharply, with the numerical results suggesting a scaling of $I_D \sim Wo^{-3}$. The suppression of $I_D$ at large $Wo$ results from the progressive localization of shear rates to thin near-wall regions (c.f. **Figure 19**) at high $Wo$, which overall reduces the shear-induced redistribution of cells across the channel.

## 4. Conclusions

We here studied the transport of a dilute suspension of elongated swimming microorganisms in oscillatory pressure-driven channel flow using complementary particle-based and continuum descriptions. We formulated a Langevin model to describe the stochastic trajectories of individual swimmers, from which we derived a two-dimensional Fokker-Planck equation governing the joint wall-normal position–orientation probability density. We first established an excellent agreement between Langevin solutions and 2D-FPE solutions over a broad range of governing parameters. We then used Langevin simulations to obtain direct physical insight into microswimmer transport under oscillatory forcing. We observed that the flow oscillations primarily decrease the magnitude of preferential concentration and shear trapping effects typically seen in steady shear flow. We attribute this to the limited timescale available for sustained shear-induced alignment within each half-cycle in oscillatory flows, which ultimately prevents the development of the orientational anisotropy required for shear trapping. Beyond this, we found that an increase in the oscillatory flow shear rate strengthens orientational alignment and enhances depletion from the channel centerline, consistent with previously reported steady flow behavior. Furthermore, similar to steady shear flows, we observed that sufficiently strong swim speeds and rotational diffusion counteract oscillatory shear-induced preferential concentration by, respectively, transporting swimmers across streamlines faster than sustained orientational alignment can develop and by randomizing swimmer orientations.

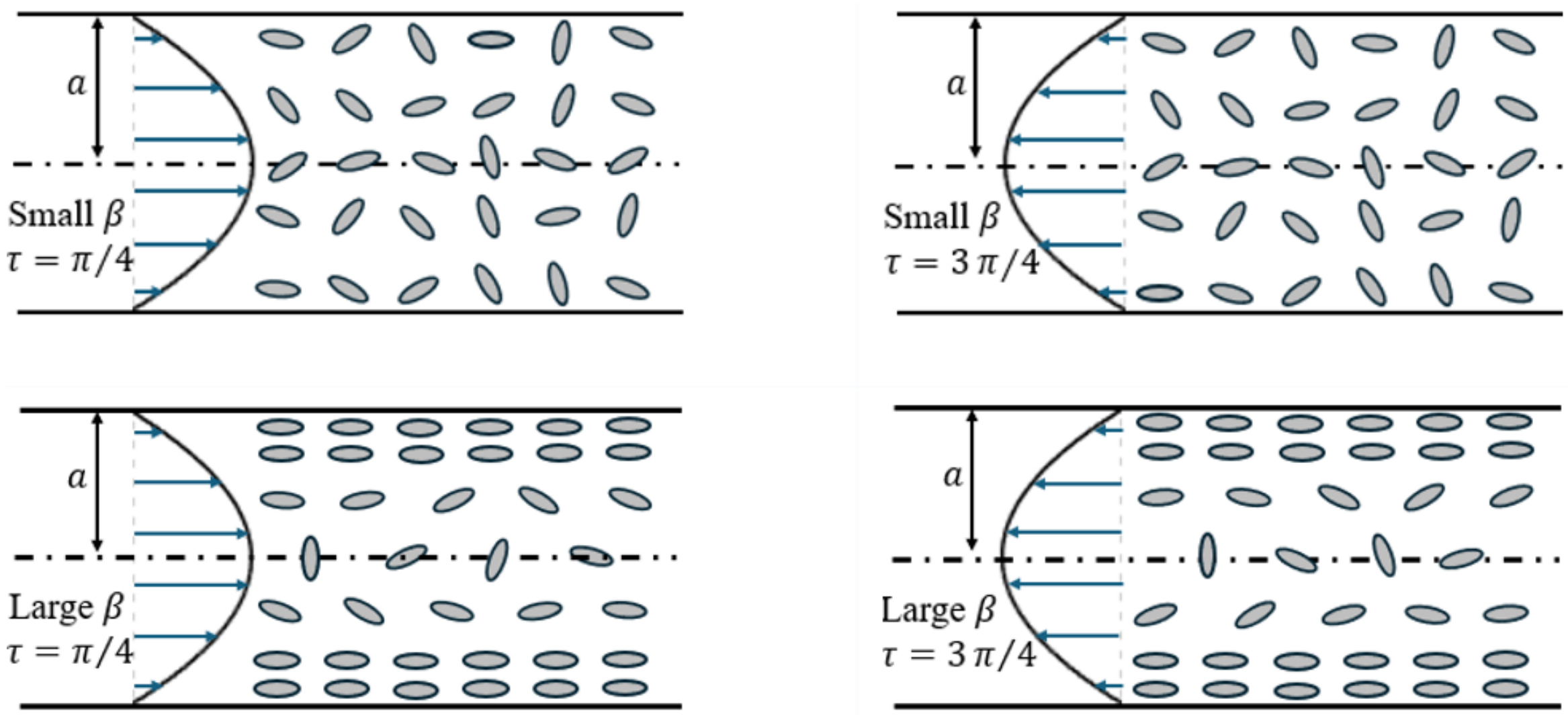

**Figure 18**. Suppression of shear trapping through temporally limited development of orientational anisotropy with decreasing frequency ratio.

We further studied the mechanisms underlying these phenomena by using a continuum model based on 2D-FPE and by systematically varying the governing dimensionless groups $Pe_f$, $Pe_c$, $Wo$ and $\beta$. The dimensionless numbers collectively characterize the competing influences of shear-induced alignment, rotational diffusion, wall-normal swimming, and oscillatory forcing on microswimmer transport. Despite the $2\pi$-periodicity of the imposed flow, we show that the microswimmer concentration field exhibits a $\pi$-periodic response. This highlights the invariance of cross-stream microswimmer redistribution to reversal of the oscillatory flow direction. Furthermore, we find that flow unsteadiness due to oscillatory flow modifies the classical steady flow shear trapping mechanism by two distinct mechanisms, which are captured by the parameters $\beta$ and $Wo$. First, an increase of the frequency ratio $\beta$, which couples swimmer and flow dynamics, exhibits a monotonically increasing but saturating influence on preferential concentration. At small $\beta$, temporal variations in the orientation distribution due to rapid flow oscillations (relative to rotational diffusion timescale) continually disrupt the build-up of the sustained anisotropy required for shear trapping (**Figure 18**). As $\beta$ increases for slower flow oscillations relative to rotational diffusion timescale, the orientational distribution achieves a quasi-steady balance between the effects of oscillatory shear-induced alignment and rotational diffusion. This leads to progressively stronger shear trapping as $\beta$ increases before the microswimmer distribution asymptotically approaches a $\beta$-independent limit. Second, an increase in the Womersley number $Wo$, which is governed solely by the imposed shear flow dynamics, weakens preferential concentration by confining the required shear rates to progressively thinner regions near the channel walls (**Figure 19**).

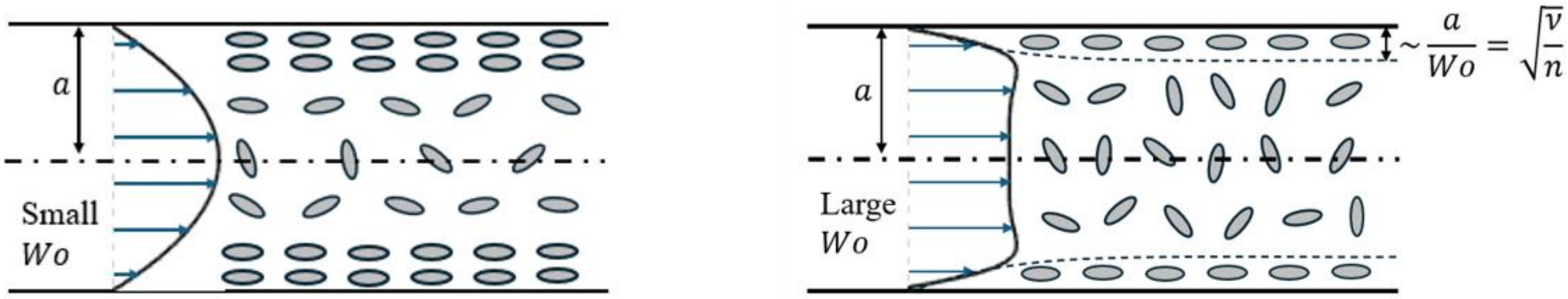

**Figure 19**. Suppression of shear trapping through spatial localization of shear with increasing Womersley number

Beyond this, we see that the effects of shear rate strength and swim speed relative to rotational diffusion, captured by $Pe_f$ and $Pe_c$, respectively, are largely similar between steady and oscillatory shear flows. Specifically, an increase in the flow Peclet number $Pe_f$ strengthens preferential concentration through enhanced shear-induced alignment. At sufficiently large values of $Pe_f$, the depletion becomes increasingly localized around the channel centerline. In the weak-swimming regime, we characterized the large-$Pe_f$ behavior by a self-similar narrowing of the depletion region accompanied by a decrease in the centerline concentration. Next, an increase in the swim Peclet number, $Pe_c$, initially strengthens preferential concentration which results in centerline depletion, but sufficiently large values of $Pe_c$ promote significant cross-stream transport which overcomes shear trapping and homogenizes the concentration field.

To gain further insight into the orientational dynamics underlying preferential concentration and the scaling behavior of microswimmer depletion index under oscillatory forcing, we developed an analytical framework based on the Fokker-Planck formulation. For this, we first derived a reduced one-dimensional Fokker-Planck equation governing the local orientational distribution and verified the solutions obtained against Langevin simulations in the limit of weak swimming. We then expressed the swimmer orientational dynamics through a Fourier harmonic representation to derive a hierarchy of coupled evolution equations for the Fourier moments, which provides a compact description of the orientational response to oscillatory shear. We obtained asymptotic solutions for the leading harmonics, which revealed a universal transfer law relating the amplitude and phase of the orientational response of microswimmers to the oscillatory forcing. Specifically, the gain and phase lag are $\alpha/\sqrt{1+16\beta^2}$ and $\tan^{-1}(1/4\beta)$. Using these results, we demonstrate that, for a given microswimmer shape, the normalized orientational response is governed solely by the frequency ratio $\beta$. We then reconstructed the orientational probability density from these solutions which provided further analytical predictions for the microswimmer concentration profile and depletion index. In particular, we show using these solutions that oscillatory flow attenuates the previously reported (Rusconi et al., 2014) quadratic dependence of the depletion index on the flow Peclet number $Pe_f$ in the weak-shear limit by the factor $16\beta^2/(1+16\beta^2)$. We showed that the analytical predictions agree well with numerical solutions over the range of validity of the perturbation analysis.

In this work, we combine stochastic particle dynamics, continuum kinetic theory, and reduced analytical modeling, to develop a unified description of microswimmer transport in oscillatory channel flows. Beyond providing quantitative predictions for preferential concentration, our work establishes a foundation for understanding active particle transport in periodically driven environments and offers a basis for the rational design of oscillatory microfluidic systems for controlling microswimmer distributions. Several extensions may follow from the present work. Future studies could address regimes of stronger swimming, for which the reduced one-dimensional description breaks down and the full coupling between wall-normal position and orientation must be retained. The two-dimensional continuum formulation could also be extended to incorporate swimmer-wall hydrodynamic interactions, finite concentration effects, and externally imposed fields. More generally, the analytical framework developed here may provide a basis for investigating transport and spatial organization of active particles in more complex time-dependent flow environments.

**Declaration of interests**
The authors report no conflict of interests.